\documentclass{aa}  

\usepackage{graphicx,txfonts}
\usepackage{natbib}
\bibpunct{(}{)}{;}{a}{}{,} 
\usepackage{hyperref}

\def\j{{(j)}}
\def\nng{n_{{\rm g}n}}

\begin{document} 

\title{The proto-neutron star inner crust in a multi-component plasma approach
\thanks{{The tables of the impurity parameter shown in Fig.~15 are available at the CDS via anonymous ftp to \url{cdsarc.u-strasbg.fr} (130.79.128.5) or via
\url{http://cdsweb.u-strasbg.fr/cgi-bin/qcat?J/A+A/}}}}

\titlerunning{The proto-neutron star inner crust in a multicomponent plasma approach}
\authorrunning{Dinh Thi et al.}

\author{H. Dinh Thi\inst{1},  
A.~F. Fantina\inst{2},
F. Gulminelli\inst{1}}
 
\institute{Université de Caen Normandie, ENSICAEN, CNRS/IN2P3, LPC Caen UMR6534, F-14000 Caen, France \\
\email{dinh@lpccaen.in2p3.fr}
\and Grand Acc\'el\'erateur National d'Ions Lourds (GANIL), CEA/DRF - CNRS/IN2P3, Boulevard Henri Becquerel, 14076 Caen, France
}
   
\date{Received xxx Accepted xxx}

  \abstract
   {Proto-neutron stars are born hot, with temperatures exceeding a few times $10^{10}$~K. In these conditions, the crust of the proto-neutron star is expected to be made of a Coulomb liquid and composed of an ensemble of different nuclear species.}
   {In this work, we perform a study of the beta-equilibrated proto-neutron-star crust in the liquid phase in a self-consistent multi-component approach. This also allows us to perform a consistent calculation of the impurity parameter, which is often taken as a free parameter in cooling simulations.}
   {To this aim, we developed a self-consistent multi-component approach at finite temperature using a compressible liquid-drop description of the ions, with surface parameters adjusted to reproduce experimental masses. 
   The treatment of the ion centre-of-mass motion was included through a translational free-energy term accounting for in-medium effects. The results of the self-consistent calculations of the multi-component plasma are systematically compared with those performed in a perturbative treatment as well as in the one-component plasma approximation.}
   {We show that the inclusion of non-linear mixing terms arising from the ion centre-of-mass motion leads to a breakdown of the ensemble equivalence between the one-component and multi-component approach. 
   Our findings also illustrate that the abundance of light nuclei becomes important and eventually dominates the whole distribution at higher density and temperature in the crust. 
   This is reflected in the impurity parameter, which, in turn, may have a potential impact on neutron-star cooling.
   For practical application to astrophysical simulations, we also provide a fitting formula for the impurity parameter in the proto-neutron-star inner crust.}
   {Our results obtained within a self-consistent multi-component approach show important differences in the prediction of the proto-neutron-star composition with respect to those obtained with a one-component approximation or a perturbative multi-component approximation, particularly in the deeper region of the crust.
   This highlights the importance of a full, self-consistent multi-component plasma calculation for reliable predictions of the proto-neutron-star crust composition.} 


   \maketitle  
%
\section{Introduction}
\label{sec:intro}

Formed from core-collapse supernova explosions, neutron stars (NSs) are initially born hot, with temperatures exceeding a few times $10^{10}\ \mathrm{K} \approx 1\ \mathrm{MeV}$ \citep{Prakash1997, Pons1999, hpy2007}.
At such temperatures, the crust of the proto-neutron star (PNS) is expected to be made of a Coulomb liquid composed of different nuclear species in a background of electrons and nucleons (see e.g. \citet{Oertel2017} for a review).
It is generally assumed that, as the NS crust cools down, this multi-component plasma (MCP) remains in full thermodynamic equilibrium until the ground state at zero temperature is reached.
Under this so-called `cold catalysed matter' hypothesis, the NS crust is considered to be made of pure layers, each consisting of a one-component Coulomb crystal.
However, if the NS cools down rapidly enough, the composition of the crust might be frozen at a temperature higher than the crystallisation temperature (see e.g. \citet{Goriely2011}), and the one-component picture becomes less reliable (see e.g. \citet{Fantina2020, Carreau2020b}).
Moreover, a correct description of the finite-temperature PNS crust requires the computation of a full nuclear ensemble.
However, due to the complexity of calculating the nuclear distribution of MCP, most approaches to modelling the finite-temperature crust have been performed within the framework of the so-called one-component plasma (OCP) approximation.
The latter assumes that, at a given thermodynamic condition, the ensemble of nuclei is represented by only one kind of nucleus, the one that is thermodynamically more favourable (see e.g. \citet{Onsi2008, Avan2009, Avan2017, Fantina2020, Carreau2020a, Carreau2020b, Dinh2022}).
This approximation is justified at relatively low densities and temperatures, where the most probable nucleus is very close to the ground-state OCP composition and the contribution of other ions is typically very small, or when computing average thermodynamic quantities \citep{Burrows1984}.
On the other hand, the coexistence of different nuclear species can have an important impact on transport properties as well as on the magneto-rotational evolution of NSs \citep{Jones2004, Pons2013} (see also \citet{Schmitt2018, Gour2018} for a review).
For these reasons, (P)NS cooling simulations are usually performed using the ground-state OCP composition, but the presence of various nuclei is taken into account via an `impurity factor', often taken as a free parameter adjusted on observational cooling data (see e.g. \citet{Vigano2013}).

Self-consistent calculations of the impurity parameter for non-accreting unmagnetised NSs at relatively low temperatures, around the crystallisation point ($k_{\rm B} T \approx 0.01 - 1$~MeV, $k_{\rm B}$ being the Boltzmann constant), were recently presented by \citet{Fantina2020} and \citet{ Carreau2020b}.
Specifically, \citet{Carreau2020b} investigated the MCP in the NS inner crust within a compressible liquid-drop (CLD) model with parameters optimised on microscopic energy-density functionals \citep{BSK24}.
 \citet{Carreau2020b} calculated the nuclear distributions using a perturbative implementation of the nuclear statistical equilibrium, as first proposed in \citet{Grams2018} for applications to supernova matter.
In this perturbative treatment, for a given baryon density $n_B$ and temperature $T$, the beta-equilibrium composition is firstly found in the OCP approximation.
This procedure yields the (OCP) neutron and proton chemical potentials, which are subsequently used in the computation of the ion abundances.
Although not fully self-consistent, this approach has the advantage of leading to a very fast convergence, with reduced computational cost with respect to a full nuclear statistical equilibrium treatment (see e.g. \citet{Botvina2010, gulrad2015, radgul2019}).
However, the applicability of such a perturbative approach requires a more thorough assessment at high densities, that is, in the deeper regions of the (P)NS crust, and at relatively high temperatures above the crystallisation temperature.

In the present work, we perform a study of the PNS crust in a self-consistent multi-component approach.
To this aim, we extended and improved the previous work of \citet{Carreau2020b} in several aspects: 
(i) we extended the study of the (P)NS inner crust to higher temperatures above the crystallisation temperature, and to higher densities, until the crust--core transition;
(ii) we improved the description of the ion centre-of-mass motion in the liquid crust by employing a more realistic prescription for the translational energy, where in-medium effects are accounted for (see the discussion in \citet{Dinh2022} and references therein);
and (iii) we went beyond the perturbative approach and performed a self-consistent calculation of the MCP in the whole PNS inner crust, which allowed us to calculate the impurity parameter in a self-consistent way.

The paper is organised as follows: The formalism is discussed in Sect.~\ref{sec:model}: the MCP approach is introduced in Sect.~\ref{sec:MCP}, the improved description of the translational free-energy term is presented in Sect.~\ref{sec:trans-MCP}, while the ion free energy in the OCP approximation is depicted in Sect.~\ref{sec:fion}, and the comparison between the MCP and OCP approaches is discussed in Sect.~\ref{sec:mostprobable}.
Numerical results are presented in Sect.~\ref{sec:results}, where we describe both the implementation of the perturbative MCP approach in Sect.~\ref{sec:approximate_MCP} and the self-consistent MCP calculations in Sect.~\ref{sec:exact_MCP}; the outcomes regarding the impurity parameter as well as a fitting formula for its practical implementation in numerical simulations are presented in Sect.~\ref{sec:Qimp}.
Finally, we draw conclusions in Sect.~\ref{sec:conlusions}.


\section{Model of the PNS inner crust}
\label{sec:model}

\subsection{MCP in nuclear statistical equilibrium}
\label{sec:MCP}

To model the full statistical distribution of ions in the PNS liquid inner crust, we extended the formalism of \citet{Carreau2020b} for its applicability at higher densities and temperatures. 
As in \citet{Carreau2020b}, we assume that beta-equilibrium holds, a condition that is met in late PNS cooling evolution stages \citep{Prakash1997, Yakovlev2004, Page2006}.
For a given thermodynamic condition defined by the total baryonic number density $n_B$ and the temperature $T$, the PNS inner crust is supposed to be composed of different spherical Wigner-Seitz (WS) cells of volume $V_{\rm WS}^\j$, each containing an ion with mass and proton number $(A^\j,Z^\j)$ and internal density $n_i^\j=A^\j/V_N^\j$ ($V_N^\j = 4 \pi r_N^{\j3} /3$ being the volume occupied by the cluster, with $r_N^\j$ the cluster radius), such that $u^\j=V_N^\j/V_{\rm WS}^\j$ is the volume fraction occupied by the cluster in the cell $j$.
Each ion is surrounded by a uniform gas of electrons and neutrons of density $n_e$ and $\nng$, respectively.
In addition, we suppose that charge neutrality is realised in every cell, meaning that the proton density is the same in each cell and equal to the electron density, that is, $n_e = n_p = Z^\j / V_{\rm WS}^\j$.

At finite temperature, a free proton gas could also be present and was included in \citet{Dinh2022}, where the PNS liquid crust was studied in the OCP approximation.
However, we find that for the considered temperature regimes, $k_{\rm B} T \lesssim 2.0$~MeV, the proton-gas density remains very small, namely about a few times $10^{-3}$~fm$^{-3}$ at most at the bottom of the crust, and its effects on the equation of state and on the crust composition are negligible.
For these reasons, we ignore the presence of the free protons in the present work, as was done in \citet{Carreau2020b}.
Moreover, in this first step to improve our previous treatment \citep{Carreau2020b}, we neglect the possible presence of non-spherical configurations, which could exist in a narrow region at the bottom of the inner crust.

Denoting $n_N^\j$, the ion density of a species $(A^\j,Z^\j)$, the frequency of occurrence or probability of the component $(j)$ is given by $n_N^\j = p_j / \langle V_{\rm WS} \rangle$, with $\sum_j p_j = 1$ and the bracket notation $\langle \rangle$ indicates ensemble averages.
The total free-energy density\footnote{We use the notation $\mathcal{F}$, upper-case $F$, and lower-case $f$ for the free-energy density, the free energy per ion, and free energy per nucleon, respectively.} of the system can therefore be written as
\begin{equation}
\mathcal{F^{\rm MCP}} = \sum_j n_N^{(j)} \left( F_i^\j -V_N^\j\mathcal{F}_{\rm g} \right) + \mathcal{F}_{\rm g} + \mathcal{F}_e  \ ,
 \label{eqMCP:FMCP}
\end{equation}
where $\mathcal{F}_{{\rm g}}$ is the free-energy density of the uniform neutron gas, see Sect. \ref{sec:fion} below, and $\mathcal{F}_e$ is the free-energy density of the uniform electron gas, for which we use the same expressions as in \citet{Carreau2020b}; see Eq.~(2.65) in \citet{hpy2007}. 
The second term in Eq.~(\ref{eqMCP:FMCP}), namely $-V_N^\j\mathcal{F}_{\rm g}$, accounts for the excluded volume, that is, the impenetrability of the different ions, 
while the cluster free energy $F_i^\j$ is given by 
\begin{equation}
F_{i}^\j = M_i^\j c^2 + F^\j_{\rm bulk} + F^\j_{\rm Coul + surf +  curv} + F^{\star,\j, {\rm MCP}}_{\rm trans} \ ,
    \label{eq:Fi0}
\end{equation}
where $M_i^\j = (A^\j-Z^\j)m_n + Z^\j m_p$ is the bare mass of the cluster, with $m_p$ ($m_n$) being the proton (neutron) mass, $c$  the speed of light, $F^\j_{\rm bulk} = \frac{A^\j}{n_i^\j} \mathcal{F}_B(n_i^\j, 1-2Z^\j/A^\j , T)$  the cluster bulk free energy, and $F^\j_{\rm Coul + surf +  curv} = F^\j_{\rm Coul} + F^\j_{\rm surf +  curv}$ is the sum of the Coulomb, surface, and curvature energies.
Finally, the last term in Eq.~(\ref{eq:Fi0}), $F^{\star,\j, {\rm MCP}}_{\rm trans}$, represents the translational free energy of the cluster in the liquid phase,
\begin{equation}
    {F}_{\rm trans}^{\star,\j, {\rm MCP}}= k_{\rm B} T \left[ \ln \left( \frac{n_N^\j}{{\Bar{u}_{\rm f}}}\frac{(\lambda_i^{\star,\j}) ^3}{g_s^\j}\right)  - 1 \right] \ ,
    \label{eq:ftrans_MCP}
\end{equation}
where $\lambda_i^{\star,\j}$ is the ion effective thermal wavelength, $g_s^\j$ is the spin degeneracy, and 
\begin{equation} 
   \Bar{u}_{\rm f} = 1- \sum_j n_N^{(j)}V_N^{(j)} \ 
    \label{eqMCP1:u}
\end{equation}
is the volume fraction available for the ion motion, 
$\Bar{u}_{\rm f} = \langle V_{\rm f} \rangle/ \langle V_{\rm WS} \rangle$,
and the average free volume is given by $\langle V_{\rm f} \rangle = \sum_j p_j V^\j_{\rm f}$, with 
\begin{equation}
    V^\j_{\rm f} = \frac{4}{3} \pi \left( r^{\j 3}_{\rm WS} - r^{\j 3}_{N} \right) \ ,
    \label{eq:VfMCP}
\end{equation}
where $r^\j_{\rm WS}$ is the WS cell radius.
The translational  term is discussed in more detail in Sect.~\ref{sec:trans-MCP}, while the bulk and interaction terms are addressed in Sect.~\ref{sec:fion}.

Following \citet{gulrad2015}, \citet{Grams2018}, \citet{Fantina2020}, and \citet{ Carreau2020b}, the densities $n_N^\j$ were calculated so as to minimise the thermodynamic potential in the canonical ensemble.
Because of the chosen free-energy decomposition in Eq.~(\ref{eqMCP:FMCP}), the electron and neutron gas free-energy densities, $\mathcal{F}_e$ and $\mathcal{F}_{\rm g}$, do not depend on $n_N^\j$.
Therefore, the variation with respect to the ion densities $n_N^\j$, $\mathrm{d} \mathcal{F}^{\rm MCP}\{n_N^\j \}/ \mathrm{d} n_N^\j$, can be performed on the ion part only, that is, on the sum term in Eq.~(\ref{eqMCP:FMCP}). 
However, the variations $\mathrm{d} n_N^\j$ are not independent because of the baryon number conservation and charge neutrality,
\begin{eqnarray}
    n_B &=&  n_{{\rm g}n} + \sum_j n_N^{(j)} \left(A^{(j)} - n_{{\rm g}n} V_N^{(j)} \right) \ , \label{eqMCP:baryon}\\
        n_e &=& n_p = \sum_j n_N^{(j)}Z^{(j)} \ . \label{eqMCP:charge}
\end{eqnarray}
These constraints are taken into account by introducing two Lagrange multipliers, which can be shown to be directly connected to the chemical potentials, yielding the following equation for the equilibrium densities $n_N^\j$:
\begin{eqnarray}
     && \sum_j \left( F_i^\j - V_N^\j \mathcal{F}_{\rm g} 
     + \sum_{j'} n^{(j')}_N \frac{\partial F_i^{(j')}}{\partial n^\j_N} \right) 
        \mathrm{d} n_N^\j \nonumber \\
      && \  + \mu_e \sum_j Z^\j \mathrm{d} n_N^\j \nonumber \\
        && \ -  \mu_{n} \sum_j \left(A^\j - \nng V_N^\j\right) \mathrm{d} n_N^\j = 0 \ .
     \label{eqMCP:dOmega_dnNj}
\end{eqnarray}
In Eq.~(\ref{eqMCP:dOmega_dnNj}), 
$\mu_e = \mathrm{d} \mathcal{F}_e / \mathrm{d} n_e$ 
is the electron chemical potential and $\mu_n$ is given by
\begin{equation}
   \mu_{n} \equiv \frac{\partial \mathcal{F}_{\rm g}}{\partial \nng}  + \Delta \mu_n = \frac{\partial \mathcal{F}_{\rm g}}{\partial \nng}  
    + \frac{\sum_j n_N^\j \left(\partial F^\j_i/ \partial \nng \right)}{1-\sum_j n_N^\j V_N^\j} \ ,
\label{eq:mun_MCP}
\end{equation}
where one can recognise the uniform-matter chemical potential of the unbound neutrons, $\mu_{{\rm g} n} \equiv \partial \mathcal{F}_{\rm g} / \partial \nng$, while the additional term $\Delta \mu_n$ accounts for the in-medium modification of the ion free energy arising from the relative motion of the ions with respect to the surrounding neutron fluid (see also \citet{Dinh2022}).
This affects the global chemical potential of the system, breaking the identity $\mu_n = \mu_{{\rm g} n}$, which is assumed in all OCP and MCP equation-of-state models based on the cluster degrees of freedom we are aware of; see \citet{Oertel2017} for a review. 
The presence of this in-medium term $\Delta \mu_n$ also implies that the pressure equilibrium condition inside each WS cell is modified with respect to the standard results, that is, the equivalence between the cluster and gas pressure (see e.g. \citet{lattimer1985}). 
In order to keep the same formal expression as in the literature ---see Eq.~(\ref{eqocp:ni}) below--- we include the in-medium modification in the definition of the gas pressure, which we define as $P_{\rm g} \equiv \mu_n \nng - \mathcal{F}_{\rm {\rm g}}$. 

The last term in the first row of Eq.~(\ref{eqMCP:dOmega_dnNj}), $\sum_{j'} n^{(j')}_N \frac{\partial F_i^{(j')}}{\partial n^\j_N}$, arises from the dependence on the ion density $n^\j_N$ of both the translational and the interaction part of the ion free energy.
Indeed, because charge conservation holds both globally ---see Eq.~(\ref{eqMCP:charge})--- and at the level of each cell, $n_e=n_p=n_p^\j = Z^\j/V_{\rm WS}^\j$, all terms depending on the cell proton density lead to a dependence on the ion abundance $n_N^\j$.
Using Eq.~(\ref{eq:ftrans_MCP}) for the translational free energy and noticing that the self-consistency is induced by the Coulomb part of the ion free energy, we obtain
\begin{equation}
    \sum_{j'} n^{(j')}_N \frac{\partial F_i^{(j')}}{\partial n^\j_N} = 
    k_{\rm B} T + \sum_{j'} n^{(j')}_N 
            \frac{\partial F_{\rm Coul}^{(j')}}{\partial n^\j_N} \ .
    \label{eq:rearr1}
\end{equation}
The last term in Eq.~(\ref{eq:rearr1}) is the so-called rearrangement term, which was shown to be important to guarantee the thermodynamic consistency of the model and to recover the ensemble equivalence between the MCP and OCP approaches (see e.g. \citet{Grams2018, Fantina2020, Carreau2020b, Barros2020}). 
Working out the rearrangement term, we find
\begin{eqnarray}
\mathcal{R}^\j &=& Z^{(j)} \sum_{j'} n_N^{(j')}
         \frac{\partial F_{\rm Coul}^{(j')}}{\partial n_p} \nonumber \\
         &=& V_{\rm WS}^\j \sum_{j'} n_N^{(j')} V_{\rm WS}^{(j')}P_{\rm int}^{(j')} \nonumber \\
         &=& V_{\rm WS}^\j \Bar{P}_{\rm int} \ ,
    \label{eq:rearr}
\end{eqnarray}
where $P_{\rm int}^{(j')}$ is the pressure due to the Coulomb interaction (see also the discussion in Appendix~\ref{sec:Ptot}),
\begin{equation}
    P_{\rm int}^{(j^\prime)} = \frac{n_p^2}{Z^{(j^\prime)}} \frac{\partial F^{(j^\prime)}_{\rm Coul}}{\partial n_p} \ .
        \label{eq:P_int}
\end{equation}
We also note that the rearrangement term, Eq.~(\ref{eq:rearr}), is not exactly equivalent to that derived in \citet{Carreau2020b}.
Indeed, in the latter work, the approximation that $F_i^\j$ was only dependent on $n_N^\j$ was made, while in reality a dependence on the other cells arises because of the Coulomb interaction, and therefore 
$\partial F_i^{(j^\prime)} / \partial n_N^\j \ne 0$ .

Considering independent variations of $n_N^\j$ and introducing the single-ion canonical potential,
\begin{equation}
    \label{eq:omegaj}
    \Omega_i^\j = F_i^{\star,\j} - V_N^\j \mathcal{F}_{\rm g} 
     + \mathcal{R}^\j - k_{\rm B} T \ln {\Bar{u}_{\rm f}} \ ,
\end{equation}
with 
\begin{equation}
    F_i^{\star,\j} = F_i^\j - {F}_{\rm trans}^{\star, (j), {\rm MCP}} 
    + k_{\rm B} T \ln \left(\frac{(\lambda_i^{\star, \j})^3}{g_s^\j}\right),
    \label{eqMCP:Fij0star}
\end{equation}
Eq.~(\ref{eqMCP:dOmega_dnNj}) becomes 
\begin{equation}
    \Omega_i^\j + k_{\rm B} T \ln n_N^\j + \mu_e Z^\j 
    -  \mu_{n} \left(A^\j - \nng V_N^\j\right) = 0 \ .
    \label{eqMCP:dOmega_dnNj-2}
\end{equation}
The ion densities can be obtained as
\begin{equation}
    n_N^\j = \exp{\left(-\frac{\tilde{\Omega}^\j_i}{k_{\rm B} T}\right)} \ ,
 \label{eqMCP:nN}
\end{equation}
with 
\begin{equation}
    \tilde{\Omega}^\j_i = \Omega_i^\j + \mu_{e} Z^\j - {\mu_{n}}  \left(A^{(j)} - \nng V_N^{(j)}  \right) \ .
 \label{eqMCP:tidle_omegaij}
\end{equation}

We note that, with respect to the derivation in \citet{Carreau2020b}, here a self-consistent problem arises.
Indeed, the terms $\Delta \mu_n$, ${\Bar{u}_{\rm f}}$, and $\Bar{P}_{\rm int}$ entering $\tilde{\Omega}^\j_i$ on the right-hand side of Eq.~(\ref{eqMCP:nN}) via Eqs.~(\ref{eq:mun_MCP}), (\ref{eqMCP1:u}), and (\ref{eq:rearr}), respectively, themselves
depend on the ion densities $n_N^\j$.
Therefore, for each given baryon density and temperature, we solve the following system of equations simultaneously:
\begin{eqnarray}
     \Delta \mu_n &=& \frac{\sum_j n_N^\j \left(\partial F^\j_i/ \partial \nng \right)}{1-\sum_j n_N^\j V_N^\j} \ , \label{eqMCP1:mun_MCP} \\
     \Bar{P}_{\rm int} &=& \sum_{j'} n_N^{(j')}  V_{\rm WS}^{(j')} P_{\rm int}^{(j')} 
     \label{eqMCP1:PMCP} \ ,
\end{eqnarray}
together with the definition of the free volume fraction, Eq.~(\ref{eqMCP1:u}), and the baryon number and charge conservation equations, Eqs.~(\ref{eqMCP:baryon}) and (\ref{eqMCP:charge}), with $n_N^{(j)}$ being given by Eq.~(\ref{eqMCP:nN}).

\subsection{Translational free energy of the ions in the MCP}
\label{sec:trans-MCP}

The translational free-energy term appearing in Eq.~(\ref{eq:Fi0}), $F^{\star,\j, {\rm MCP}}_{\rm trans}$, accounts for the centre-of-mass motion of each ion of the liquid MCP.
Contrarily to the OCP description of the crust (see \citet{Dinh2022} for a detailed discussion of the translational term and its impact on the OCP inner-crust composition and equation of state), in the MCP picture, the centre-of-mass position of each ion is not confined to the single WS cell; the cluster can explore the whole volume, albeit with an effective mass that accounts for the fact that only bound nucleons can be in relative motion with respect to the unbound neutron gas. 
This leads to Eq.~(\ref{eq:ftrans_MCP}) for the translational free energy, where we set the spin degeneracy $g_s^\j = 1$, because the most abundant nuclei are expected to be even-even. 
The effective thermal wavelength of the ion in Eq.~(\ref{eq:ftrans_MCP}) is given by
\begin{equation}
    \lambda^{\star,\j}_i = \sqrt{\frac{2\pi \hbar^2}{M^{\star,\j}_i k_{\rm B} T}},
    \label{eq:lambda_i}
\end{equation}
with $\hbar$ being the Planck constant, and the effective mass reads 
\begin{equation}
 M^{\star,\j}_i = M^\j_i (1-\gamma^\j) \ ,
\label{eq:Mstar_i}
\end{equation}
with $\gamma^\j = \nng/n^\j_i$ being the ratio between the neutron-gas density and the cluster density.
We note that Eq.~(\ref{eq:ftrans_MCP}) differs from the translational free energy used in \citet{Carreau2020b} in two respects:
(i) here we use a different expression for the effective mass, that is, the one derived in the hydrodynamic approach considering that only neutrons in the continuum participate in the (collective) flow (see also Sect.~2 in \citet{Dinh2022}); and
(ii) the centre-of-mass position of each ion of the MCP cannot freely explore the whole volume, which is consistent with the excluded-volume approach of Eq.~(\ref{eqMCP:FMCP}).
In accordance with the approach of \citet{Hempel2010}, the latter condition gives rise to the extra ${\Bar{u}_{\rm f}}$ term in Eq.~(\ref{eq:ftrans_MCP}).  
 
 \citet{Fantina2020} noted that, because of the different expressions of the translational energy in the OCP and MCP approaches, a first deviation from the linear mixing rule appears. 
As a consequence, the total 
free energy of the ions is not just the sum of the OCP ion free energies, but an extra term arises, known in the literature as the mixing entropy (see Eq.~(21) in \citet{Fantina2020}; see also \citet{Medin2010}).
This is still the case here, but the additional (non-ideal) `mixing entropy' term now reflects the in-medium (excluded-volume) effects included in the translational energy $F_{\rm trans}^{\star,\j, {\rm MCP}}$,
\begin{equation}
\label{eq:Fi0-2}
F_{i}^\j = F_i^{\j, {\rm OCP}} + k_{\rm B}T \ln \left[ p_j \frac{Z^\j}{\langle Z \rangle} \frac{\left( 1-(u^\j)^{1/3} \right)^3}{{\Bar{u}_{\rm f}}}\right] \ .
\end{equation}
The computation of the free energy of the ions in the MCP therefore still requires knowledge of the ion free energy in the OCP phase, which is addressed in Sect.~\ref{sec:fion}.


\subsection{Ion free energy in the OCP approximation}
\label{sec:fion}

We define in this section the different terms of the ion free energy in the OCP approximation, $F_i^{\j, {\rm OCP}}$; see Eq.~(\ref{eq:Fi0-2})\footnote{In this section, to simplify the notation, we drop the $\j$ superscript.}.
A detailed discussion on these terms can also be found in \citet{Dinh2022}; here we present the main expressions, for completeness.

In the CLD approach, the ion free energy is decomposed in the bulk, Coulomb, finite-size, and translational contributions (see also Eq.~(\ref{eq:Fi0})), 
\begin{equation}
F_{i}^{\j, {\rm OCP}} = F_i = M_i c^2 + F_{\rm bulk} + F_{\rm Coul + surf +  curv} + F^{\star}_{\rm trans} \ .
    \label{eq:Fi-OCP}
\end{equation}
The bulk free energy of the cluster, $F_{\rm bulk} = \frac{A}{n_i} \mathcal{F}_B(n_i, 1-2Z/A , T)$, is expressed in terms of the free-energy density of homogeneous nuclear matter $\mathcal{F}_B(n,\delta,T)$ at a total baryonic density $n = n_n + n_p$ ($n_n$ and $n_p$ being the neutron and proton densities, respectively), isospin asymmetry $\delta = (n_n - n_p)/n$, and temperature $T$. 
We note that the same expression of the bulk energy is used to compute the free energy of the free neutron gas in Eq.~(\ref{eqMCP:FMCP}), $\mathcal{F}_{\rm g}=\mathcal{F}_B(n=\nng, \delta_{\rm g} =1 , T) + \nng m_n c^2$. 
To determine $\mathcal{F}_B$, we used the self-consistent mean-field thermodynamics \citep{lattimer1985, Ducoin2007} (see Sect.~3.1 in \citet{Dinh2022} for details), where the free-energy density is decomposed into a `potential' and a `kinetic'     part,
\begin{equation}
\mathcal{F}_B(n,\delta,T) = \mathcal{F}_{\rm kin}(n,\delta,T) + \mathcal{V}_{\rm MM}(n,\delta) \ .
\label{eq:Fb}
\end{equation}
The `kinetic' term $\mathcal{F}_{\rm kin}$ in Eq.~(\ref{eq:Fb}) is given by
\begin{equation}
\mathcal{F}_{\rm kin} = \sum_{q=n,p} \left[ \frac{-2 k_{\rm B} T}{\lambda_q^{3/2}} F_{3/2}\left( \frac{\tilde{\mu}_q}{k_{\rm B} T} \right) + n_q \tilde{\mu}_q  \right] \ ,
\label{eq:Fkin}
\end{equation}
where $q=n,p$ labels neutrons and protons, $\lambda_q = \left( \frac{2 \pi \hbar^2}{k_{\rm B} T m^\star_q} \right)^{1/2}$ is the thermal wavelength of the nucleon with $m^\star_q$ being the density-dependent nucleon effective mass, $F_{3/2}$ denotes the Fermi-Dirac integral, and the effective chemical potential $\tilde{\mu}_q$ is related to the thermodynamical chemical potential $\mu_q$ through $\tilde{\mu}_q = \mu_q - U_q$, with $U_q$ the mean-field potential.
Regarding the `potential' energy, we employed the meta-model approach of \citet{Margueron2018a, Margueron2018b}, where $\mathcal{V}_{\rm MM}(n,\delta)$ is expressed as a Taylor expansion truncated at order $N$ (here, $N=4$) around the saturation point ($n=n_{\rm sat}$, $\delta=0$),
\begin{equation}
   \mathcal{V}^{N = 4}_{\rm MM}(n, \delta) = \sum_{k=0}^{4} \frac{n}{k!}(v^{\rm is}_{k} +v^{\rm iv}_{k}\delta^2 )x^{k}u^{N = 4}_{k}(x),
   \label{eq:vMM}
\end{equation}
where $x = (n - n_{\rm sat})/(3n_{\rm sat})$ and $u^{N}_{k}(x) = 1 - (-3x)^{N+1-k}\exp(-b(1+3x))$, with $b = 10\ln(2)$ being a parameter governing the function behaviour at low densities. 
The factor $u^{N}_{k}(x)$ was introduced to ensure the convergence at the zero-density limit, while the parameters $v^{\rm is}_{k}$ and $v^{\rm iv}_{k}$ are linear combinations of the so-called nuclear matter empirical parameters $\{ n_{\rm sat},E_{\rm sat,sym},L_{\rm sym},K_{\rm sat,sym},Q_{\rm sat,sym},Z_{\rm sat,sym}\}$ (see \citet{Margueron2018a} for details). 
In this work, we employ the BSk24 empirical parameters \citep{BSK24}, as an illustrative example; indeed, this functional is in very good agreement with all current astrophysical data as well as nuclear physics experiments and theory \citep{Dinh2021a}.  

The Coulomb and finite-size contributions, ${F}_{\rm Coul + surf +  curv} = V_{\rm WS} (\mathcal{F}_{\rm Coul} + \mathcal{F}_{\rm surf + curv})$, account for the total interface energy.
As we consider here only spherical clusters, the Coulomb energy density is given by 
\begin{equation}
    \mathcal{F}_{\rm Coul}  = \frac{2}{5}\pi (e n_i r_N)^2 u \left(\frac{1-I}{2}\right)^2 \left[ u+ 2  \left( 1- \frac{3}{2}u^{1/3} \right) \right] \ , 
    \label{eq:Fcoul}
\end{equation}
with $e$ being the elementary charge and $I = 1-2Z/A$.
For the surface and curvature contributions, we employed the same expression as in \citet{Lattimer1991}, \citet{Maru2005}, and \citet{Newton2013}, that is
\begin{equation}
\mathcal{F}_{\rm {surf}} + \mathcal{F}_{\rm {curv}} =\frac{3u}{r_N} \left( \sigma_{\rm s}(I, T) +\frac{2\sigma_{\rm c}(I, T)}{r_N} \right) \ , 
\label{eq:interface}   
\end{equation}
where $\sigma_{\rm s}$ and $\sigma_{\rm c}$ are the surface and curvature tensions \citep{Lattimer1991},
\begin{equation}
\sigma_{\rm s(c)}(I,T) = \sigma_{\rm s(c)}(I, T=0) h(T) \ .
\label{eq:sigmaT}
\end{equation}
The expressions of the surface and curvature tensions at zero temperature are taken from \citet{Ravenhall1983}, who proposed a parameterization based on the Thomas-Fermi calculations at extreme isospin asymmetries,
\begin{eqnarray}
\sigma_{\rm s}(I, T=0) &=& \sigma_0 \frac{2^{p+1} + b_{\rm s}}{y_p^{-p} + b_{\rm s} + (1-y_p)^{-p}} \ , \\
\sigma_{\rm c} (I,T=0) &=& 5.5 \sigma_{\rm s}(I,T=0) \frac{\sigma_{0, {\rm c}}}{\sigma_0} (\beta -y_p) \ ,
\label{eq:sigma0}
\end{eqnarray}
where $y_p = (1-I)/2$ and the surface parameters $(\sigma_0, \sigma_{0, {\rm c}}, b_{\rm s}, \beta,p)$ were optimised for each set of bulk parameters and effective mass to reproduce the experimental nuclear masses in the 2016 Atomic Mass Evaluation (AME) table \citep{AME2016}. 
The temperature dependence of the surface terms is encoded in the function $h(T)$ \citep{Lattimer1991}
\begin{equation}
h(T) = 
\begin{cases} 
  0 & \text{if $T > T_{\rm c}$}  \\
  \left[1 -\left(\frac{T}{T_{\rm c}}\right)^2\right]^2 & \text{if $T \leq T_{\rm c}$}
 \end{cases} \ ,
\end{equation}
where $T_{\rm c}$ is the critical temperature (see Eq.~(2.31) in \citet{Lattimer1991}).

Finally, the translational energy in the OCP approximation is given by \citep{Dinh2022}
\begin{equation}
    {F}_{\rm trans}^\star= k_{\rm B}T\ln \left( \frac{1}{{V^{\rm OCP}_{\rm f}}} \frac{\lambda_i^{\star 3}}{g_s} \right)  - k_{\rm B}T \ , 
    \label{eq:ftrans_eff}
\end{equation}
where the reduced (`free') WS volume reads \citep{lattimer1985}
\begin{equation}
    {V^{\rm OCP}_{\rm f}} = \frac{4}{3} \pi (r_{\rm WS} - r_{N})^3 \ .
    \label{eq:Vf}
\end{equation}
We remark that the definition of the reduced volume, Eq.~(\ref{eq:Vf}), differs from that introduced 
in Sect.~\ref{sec:MCP} for the MCP translational free energy, Eq.~(\ref{eq:VfMCP}).
Indeed, in the MCP picture, the WS approximation is relaxed and the integral over the cluster centre of mass is extended to the whole (macroscopic) volume, except that occupied by the ensemble of the different ions, which leads to Eqs.~(\ref{eq:ftrans_MCP}) and (\ref{eq:VfMCP}).

\subsection{MCP versus OCP}
\label{sec:mostprobable}

In order to compare the MCP and OCP approaches, we evaluated the most probable configuration in the MCP.
The latter corresponds to the maximum probability, $p_j^{\rm max}$, or equivalently the minimum grand-canonical potential, $\tilde{\Omega}^{(j), \rm min}_i$. 
The minimisation was performed with respect to the three variables associated to the cluster, that is ($r_N^\j,I^\j,n_i^\j)$, yielding the following system of equations:
 
\begin{eqnarray}
 && (n_i^\j)^2  \frac{\partial (F_i^{\star,\j}/A^\j)}{\partial n_i^\j} -  \frac{r_N^\j  n_i^\j}{3} \frac{\partial (F_i^{\j, \star}/A^\j)}{\partial r_N^\j}  \nonumber \\
 && \ \ \ \ \ \ \ = P_{\rm g} , \label{eqMCP:mp1} \\
&&  \frac{r_N^\j}{3A^\j} \frac{\partial F_i^{\star,\j}}{\partial r_N^\j} + (1 - I^\j) \frac{\partial (F_i^{\star,\j}/A^\j)}{\partial I^\j} \nonumber \\
&& \ \ \ \ \ \ \ = {\mu_{n}}  - \frac{P_{\rm g}}{n_i^\j}, \label{eqMCP:mp2} \\
&& 2 \frac{\partial (F_i^{\star,\j}/A^\j)}{\partial I^\j} - \frac{\Bar{P}_{\rm int}}{n_p} = \mu_e \ . \label{eqMCP:mp3}
\end{eqnarray}
 
This system of variational equations can be compared to that obtained in the OCP approach of \citet{Carreau2020b} and \citet{Dinh2022}, without the free proton gas,
\begin{eqnarray}
  n_i^2\frac{\partial}{\partial n_i}\left(\frac{F_i}{A}\right) &=&  P_{\rm g} \ , \label{eqocp:ni}\\
         \frac{F_i}{A} + (1-I) \frac{\partial}{\partial I}\left(\frac{F_i}{A}\right) &=&  \mu_n - \frac{P_{\rm g}}{n_i} \ ,  \label{eqocp:I}\\
  2\left[  \frac{\partial}{\partial I}\left(\frac{F_i}{A}\right) - \frac{n_p}{1-I}  \frac{\partial}{\partial n_p}\left(\frac{F_i}{A}\right) \right] &=& \mu_e \ , \label{eqocp:Iandnp} \\
         \frac{\partial (F_i/A)}{\partial r_N} &=& 0 \ . \label{eqocp:rn}
\end{eqnarray}
It is easy to show that Eqs.~(\ref{eqMCP:mp1})-(\ref{eqMCP:mp3}) 
for the most probable ion in the MCP are equivalent to the first three OCP variational equations Eqs.~(\ref{eqocp:ni})-(\ref{eqocp:Iandnp})
if the following conditions are satisfied: 
(i) $F_i^{\star, \j}$ is the same as in the OCP,  that is,   $F_i^{\star, \j} = F_i^{\j, {\rm OCP}} = F_i$ (see Sect.~\ref{sec:fion}; this is the case if the non-linear mixing term, which arises due to the translational motion \citep{gulrad2015, Fantina2020}, see Sect.~\ref{sec:trans-MCP}, is negligible); and 
(ii) the gas densities, $\nng$ and $n_e$, as well as $\Delta \mu_n$ and $\Bar{P}_{\rm int}$, are identical to the OCP values.
This latter condition is expected to be satisfied if the ion distributions are strongly peaked on a unique ion species, such that the averages in Eqs.~(\ref{eqMCP1:mun_MCP})-(\ref{eqMCP1:PMCP}) can be approximated to a single term corresponding to the OCP solution.
The validity of these conditions, which are necessary for the OCP approximation to give a satisfactory description of the finite-temperature configuration, are discussed in Sect.~\ref{sec:approximate_MCP}.

The additional OCP condition in Eq.~(\ref{eqocp:rn}) is identical to the well-known Baym virial theorem \citep{bbp}, which holds for a nucleus in vacuum. 
This condition arises from the fact that, for a given thermodynamic condition, there is systematically one more independent variable in the OCP than in the MCP. Indeed, the WS cell volume in the MCP is simply defined by charge conservation, while in the OCP it corresponds to an additional variable which has to be variationally determined. 

In a CLD picture, for a given mass number $A$ and atomic number $Z$, a nuclear species is also characterised by its radius, $r_N$, or equivalently its density $n_i=3A/(4\pi r_N^3)$, which can in principle fluctuate from cell to cell in the MCP equilibrium. 
Changing variables from $(r_N,I,n_i)$ to $(A,Z,n_i)$, the ion-radius distribution of a given $(A,Z)$ nucleus is expressed as
\begin{equation} 
    p_{{A}{Z}}(r_N)=\mathcal{N}n_N(A^\j= A, Z^\j= Z,r_N),
\end{equation}
where $\mathcal{N}$ is a normalization. 
This distribution will be peaked at a value $r_N$ corresponding to
 ${\partial\tilde{\Omega}^\j_i  }/{\partial n_i}|_{A^\j,Z^\j}  = 0$, 
 yielding the following pressure equilibrium condition:
\begin{equation} \label{eq:Peq}
P^{\star}_{\rm cl } \equiv \left.\frac{n_i^{2}}{A} \frac{\partial F_i^{\star} }{\partial n_i} \right|_{A,Z} = P_{\rm g}.
\end{equation}
Equation~(\ref{eqocp:ni}) is not necessarily compatible with the so-called virial condition, or radius-minimisation ansatz given by 
 \begin{equation} \label{eq:req}
  \left.\frac{\partial (F_i^{\star}/A)}{\partial r_N}\right|_{A,Z} = 0 \ . 
 \end{equation}
 As a consequence, even if conditions (i) and (ii) above are met, the minimisation equation for the OCP radius, Eq.~(\ref{eqocp:rn}), is not necessarily fulfilled in the MCP. 

\begin{figure}
    \centering
    \includegraphics[scale = 0.4]{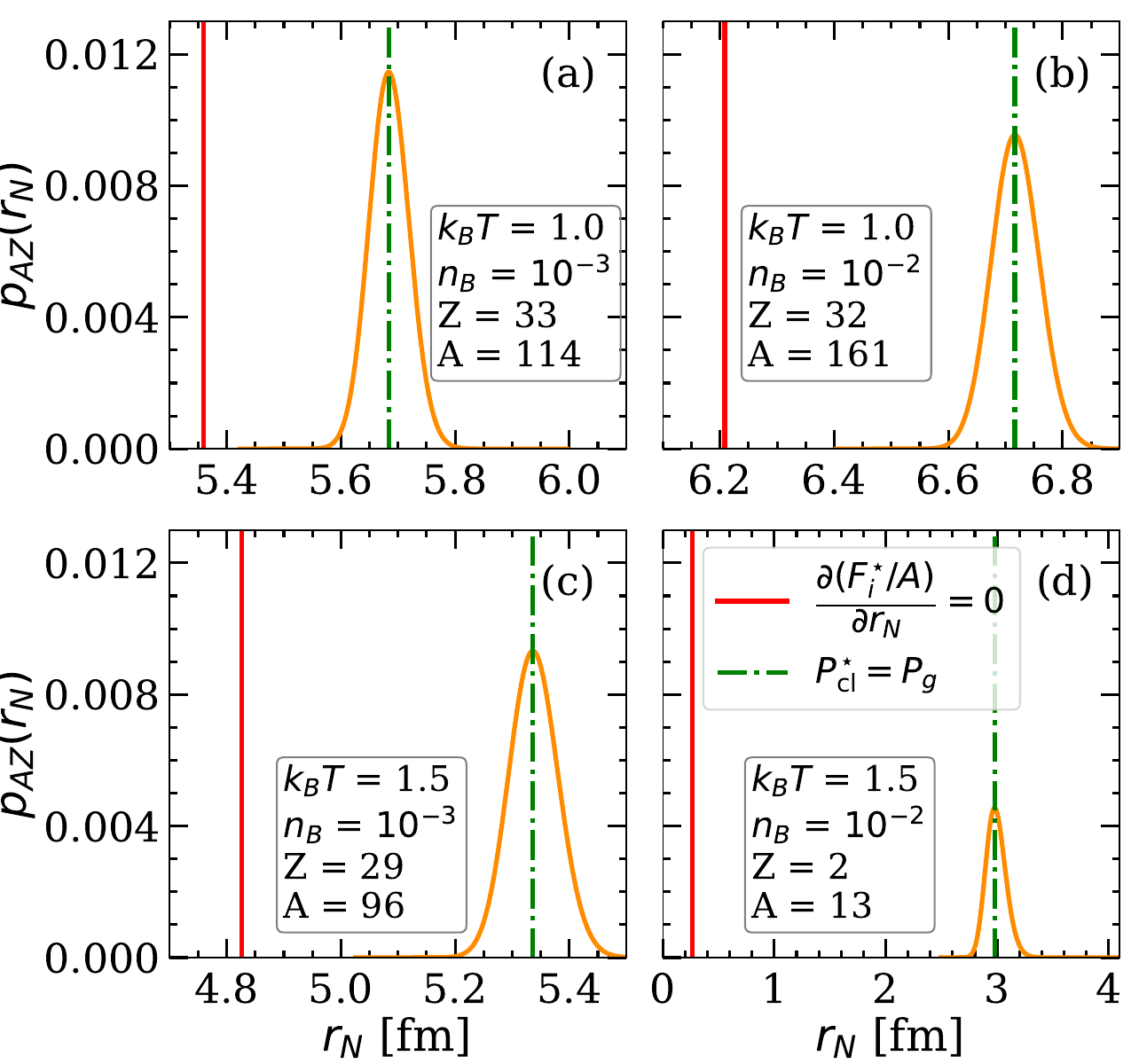}
    \caption{Normalised distribution of the cluster radius, $p_{AZ}(r_N)$ (orange solid line), in a full MCP calculation. For comparison, the solutions from two different equilibrium prescriptions for the most probable cluster are displayed. Results are presented for two different temperatures: $k_{\rm B} T = 1$~MeV (top panels) and $k_{\rm B} T = 1.5$~MeV (bottom panels), and two different densities: $n_B = 10^{-3}$~fm$^{-3}$ (left panels) and $n_B = 10^{-2}$~fm$^{-3}$ (right panels). See text for details.}
    \label{fig:2loops_vs_3loops}
\end{figure}

This first important difference between the OCP and MCP approaches is presented in Fig.~\ref{fig:2loops_vs_3loops}, where we plot the normalised distribution of the cluster radius for the most probable $(A,Z)$ nucleus obtained at different thermodynamic conditions (orange solid lines).
This distribution is compared with the solution obtained from the condition of pressure equilibrium in Eq.~(\ref{eq:Peq}) (green dash-dotted lines) and that resulting from the minimisation of the free energy per nucleon with respect to $r_N$ in Eq.~(\ref{eq:req}) (red solid lines).
To obtain these distributions, at each thermodynamic condition defined by ($n_B$, $T$), we considered independent variations of the three variables $(A^\j, Z^\j, r_N^\j)$, and solved the system of equations given by Eqs.~(\ref{eqMCP:baryon}), (\ref{eqMCP:charge}), (\ref{eqMCP1:u}), (\ref{eqMCP1:mun_MCP}), and (\ref{eqMCP1:PMCP}), together with the equation for the ion density $n_N^\j$, Eq.~(\ref{eqMCP:nN}) (see also Sect.~\ref{sec:exact_MCP}).
From Fig.~\ref{fig:2loops_vs_3loops}, one can also observe that the solution of the pressure equilibrium equation always coincides with the peak of the MCP distribution. 
This is true not only for the radius of the most probable cluster, but also for the $r_N^\j$ of any $(A^\j, Z^\j)$ cluster. 
On the other hand, using Eq.~(\ref{eq:req}) to determine the cluster radius fails in reproducing the most probable cluster radius obtained within the MCP approach.
In fact, it only produces the correct peak for the most probable cluster if the OCP chemical potentials are used in the absence of the non-linear mixing term, as it was analytically demonstrated in \citet{Grams2018}.

Computation of the crust observables considering the cluster distribution in the three-dimensional variable space $(A,Z,r_N)$ is a heavy task from a numerical point of view. 
In the literature, most MCP codes designed to determine the composition and equation of state for core-collapse simulations or for PNS modelling either ignore the cluster density degree of freedom (see e.g. \citet{Hempel2010, radgul2019}) or impose the virial condition, Eq.~(\ref{eq:req}) (see e.g. \citet{Furusawa2017,Barros2020, Pelicer2021}).
To limit the numerical cost, in the following sections we only consider a two-dimensional cluster distribution $p_{AZ}$, and fix for each $(A,Z)$ the cluster radius $r_N$ from the pressure equilibrium condition using Eq.~(\ref{eq:Peq}).
Indeed, we verified that, in all thermodynamic conditions explored in the present work, the radius distribution is always symmetric and relatively narrow.


\section{Crust composition at finite temperature in the MCP} \label{sec:results}

Using the formalism described in Sect.~\ref{sec:MCP}, we calculated the composition and equation of state of the inner crust of a PNS, employing the  
BSk24 \citep{BSK24} 
empirical parameters.
We performed our analysis for temperatures in the range from 1~MeV to 2~MeV, where the crust is expected to be in the liquid phase \citep{Carreau2020a} and the beta-equilibrium condition be realised.

\subsection{Perturbative MCP calculations}
\label{sec:approximate_MCP}

We begin this section by discussing  the results obtained by calculating the MCP distribution in a perturbative approach, as first proposed in \citet{Grams2018}.
For each $(n_B, T)$, we first calculated the OCP solution, solving the system of equilibrium equations Eqs.~(\ref{eqocp:ni})-(\ref{eqocp:rn}). 
This yields the OCP composition (i.e. $A^{\rm OCP}, Z^{\rm OCP}$, and the nuclear radius $r_N^{\rm OCP}$), the OCP chemical potentials, and the neutron and electron densities, $\nng = \nng^{\rm OCP}$, $n_{e} = n_e^{\rm OCP}$.
With the latter five quantities as input, and 
 \begin{eqnarray}
      \Delta \mu_n &\equiv & \Delta \mu_n^{\rm OCP} = \left(\frac{\partial F_{i}/\partial \nng}{{V_{\rm WS} - V_N}}\right)_{\rm OCP} \ , \label{eq:demun_atOCP} \\
       \Bar{P}_{\rm int} &\equiv& \Bar{P}_{\rm int}^{\rm OCP} = \left(\frac{n_p^2}{Z} \frac{\partial F_{\rm Coul}}{\partial n_p} \right)_{\rm OCP} \ , \\
       \bar{u}_{\rm f} &\equiv& \bar{u}_{\rm f}^{\rm OCP} = \left( \frac{V_{\rm f}}{V_{\rm WS}} \right)_{\rm OCP}  \ ,
    \end{eqnarray}
the MCP calculations were performed, giving the normalised probabilities $p_j\equiv p_{AZ}$.

\begin{figure}
    \centering
    \includegraphics[scale = 0.4]{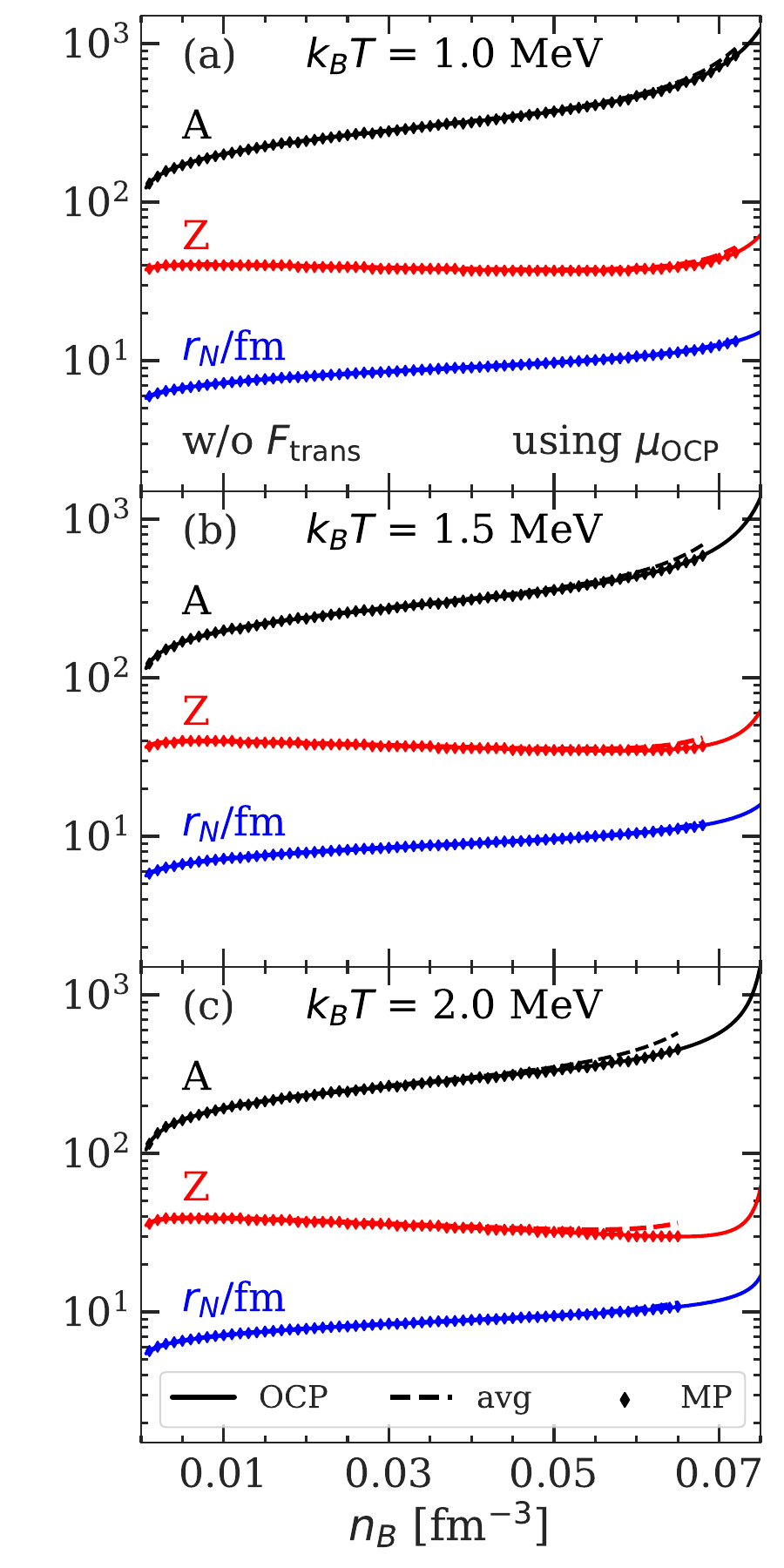}
    \caption{Cluster radius $r_N$ (blue), proton number $Z$ (red), and mass number $A$ (black) as a function of the  baryon density $n_B$ at different temperatures: $k_{\rm B}T = 1.0$~MeV (panel a), $k_{\rm B}T = 1.5$~MeV (panel b), and $k_{\rm B}T = 2.0$~MeV (panel c). OCP results are shown by solid lines, while dashed lines and symbols correspond to the average (avg) and the most probable (MP) value, respectively, of the cluster distribution in a perturbative MCP calculation. 
    The translational free energy is not included in the calculations. }
    \label{fig:ppMCP_focp_mcpvsocp_without_ftrans}
\end{figure}

As discussed in Sect.~\ref{sec:mostprobable}, the OCP and MCP results should coincide if, in addition, the non-linear mixing term coming from the translational free energy is set to zero.
This is illustrated in Fig.~\ref{fig:ppMCP_focp_mcpvsocp_without_ftrans}, where the cluster radius $r_N$ (blue curves), proton number $Z$ (red curves), and mass number $A$ (black curves) predicted by the OCP (solid lines) are plotted as a function of the input baryon density, $n_B$, together with the MCP average (dashed lines) and most probable quantities (diamonds).
The latter values correspond to those maximising $p_{AZ}$.
For all the considered densities (from the neutron-drip up to the crust--core transition density) and temperatures, we can see that the most probable $A$, $Z$, and $r_N$ coincide with the OCP solutions, and follow the average values very closely, implying that the distributions are centred around the OCP predictions. 

\begin{figure}
    \centering
    \vspace{-0.5cm}
    \includegraphics[scale = 0.4]{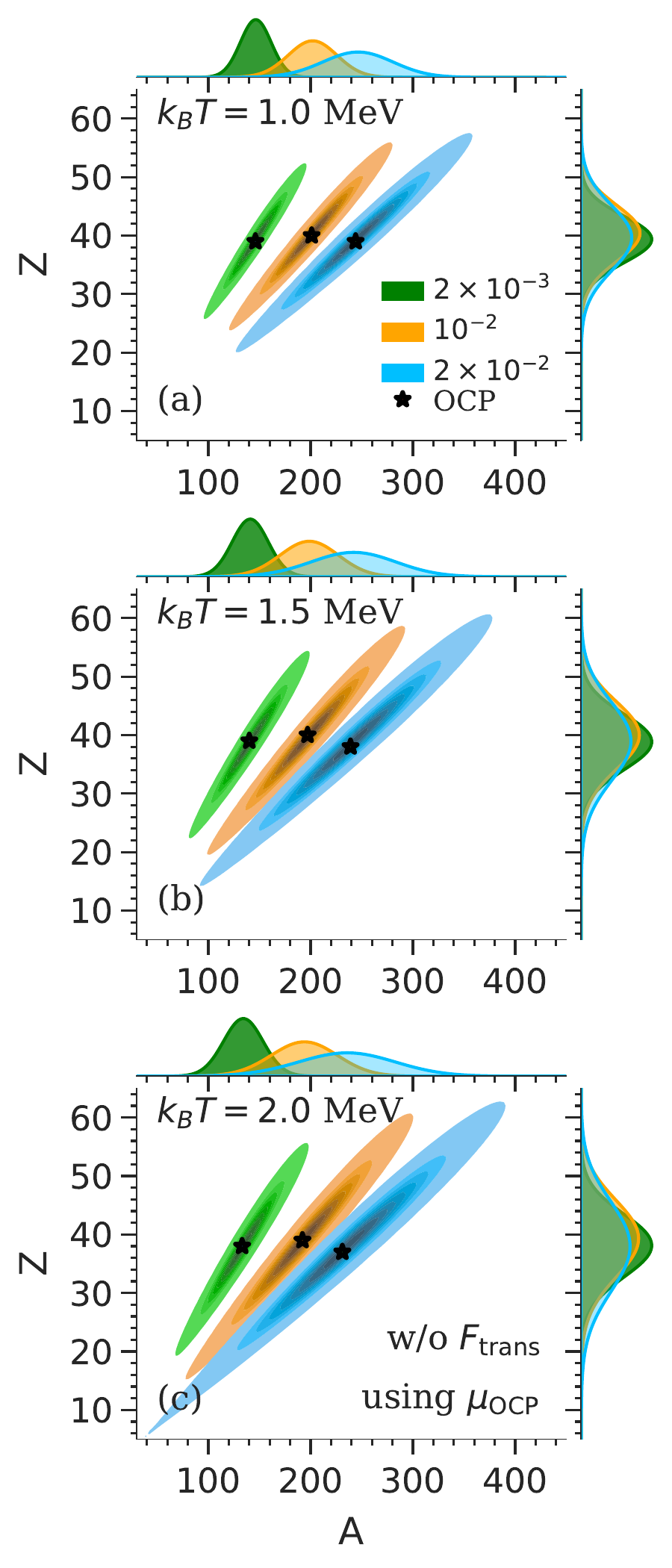}
    \caption{Joint distributions of the cluster proton number $Z$ and mass number $A$ for the same temperatures as in Fig.~\ref{fig:ppMCP_focp_mcpvsocp_without_ftrans} and for three selected input baryon densities: $n_B=2\times 10^{-3}$~fm$^{-3}$ (green contours), $n_B =  10^{-2}$~fm$^{-3}$ (orange contours), and $n_B = 2\times 10^{-2}$~fm$^{-3}$ (blue contours), in a perturbative MCP calculation. The black stars indicate the OCP solution. The translational free energy is not included in the calculations.}
    \label{fig:ppMCP_focp_distributions_woFtrans}
\end{figure}

This is further shown in Fig.~\ref{fig:ppMCP_focp_distributions_woFtrans}, where we display the joint distributions of the cluster proton number $Z$ and mass number $A$ for the same temperatures as in Fig.~\ref{fig:ppMCP_focp_mcpvsocp_without_ftrans} and for three selected densities, $n_B  = 2\times 10^{-3}$~fm$^{-3}$ (green contours), $n_B  =  10^{-2}$~fm$^{-3}$ (orange contours), and $n_B  = 2\times 10^{-2}$~fm$^{-3}$ (blue contours).
The distributions are indeed Gaussian-like, peaked at the OCP solutions (black stars).
We can also observe that, while the proton number $Z$ is relatively constant, the most probable mass number $A$ increases with density, as already noted in \citet{Carreau2020b} and \citet{Dinh2022}.
Moreover, the width of the distributions gets broader with temperature and density, with lighter clusters being populated at higher $T$ and $n_B$ (see panel c), underlying the importance of considering a full nuclear ensemble instead of a single-nucleus (OCP) approach.

\begin{figure}
    \centering
    \includegraphics[scale = 0.4]{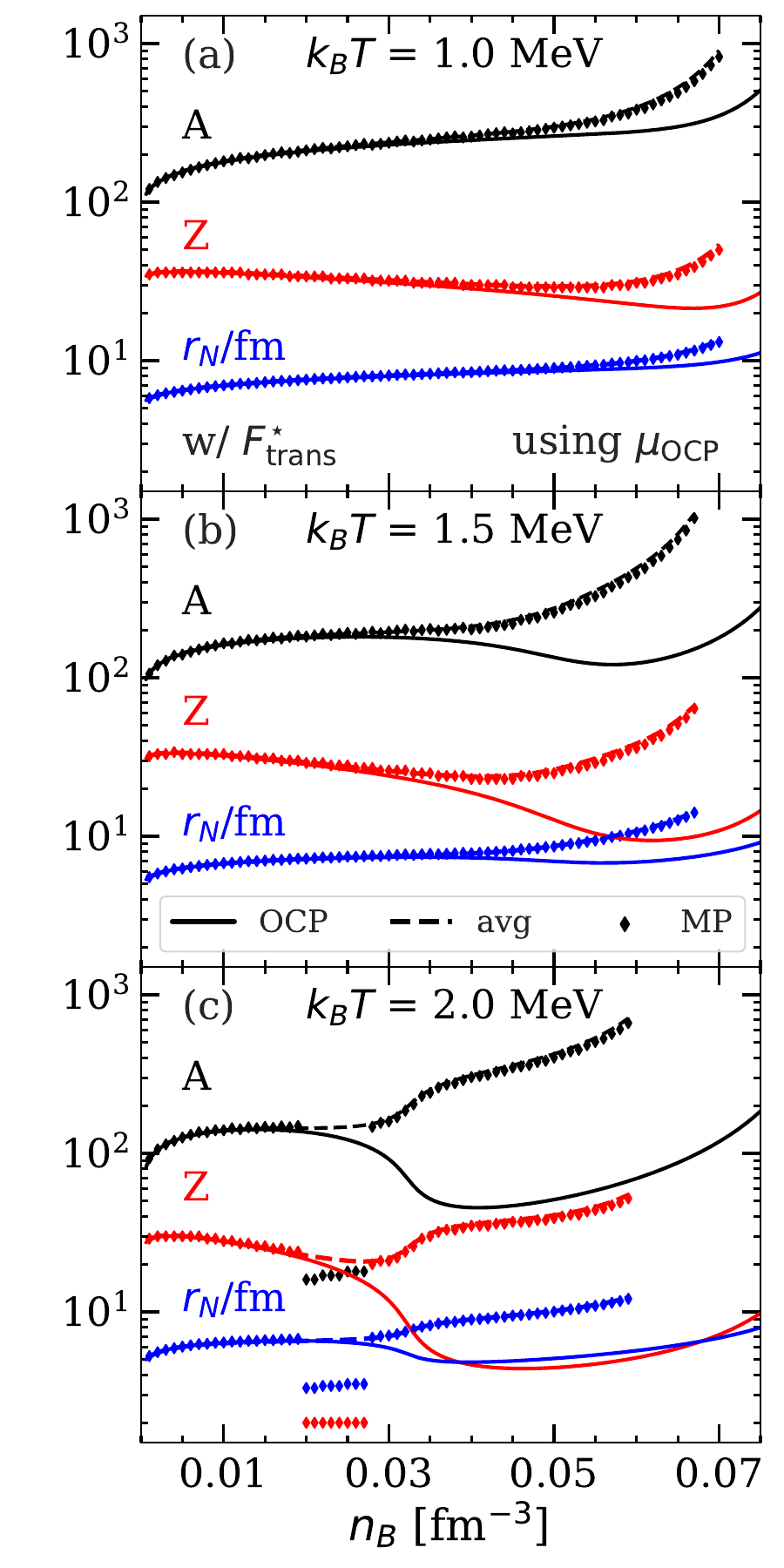}
    \caption{Same as in Fig.~\ref{fig:ppMCP_focp_mcpvsocp_without_ftrans} but with the translational free energy, $F^{\star}_{\rm trans}$, included in both the OCP and perturbative MCP calculations. }
    \label{fig:ppMCP_focp_mcpvsocp_with_modified_ftrans}
\end{figure}

We now turn to a discussion of the influence of the non-linear mixing term, which comes from the translational motion. 
In Fig.~\ref{fig:ppMCP_focp_mcpvsocp_with_modified_ftrans}, we show the evolution with the baryon density of the cluster variables $A$, $Z$, and $r_N$, for the same conditions as in Fig.~\ref{fig:ppMCP_focp_mcpvsocp_without_ftrans}, but with the translational free energy included in both the OCP and the MCP calculations, Eqs.~(\ref{eq:ftrans_eff}) and (\ref{eq:ftrans_MCP}), respectively. 
We can see that the different description of the centre-of-mass motion induces a discrepancy between the predictions of the two approaches.
Indeed, in the OCP approximation, the cluster moves in the reduced (`free') volume $V^{\rm OCP}_{\rm f}$ ---see Eqs.~(\ref{eq:ftrans_eff})-(\ref{eq:Vf})--- associated with the single WS cell determined from the variational procedure, while in the MCP, clusters of different species are considered to move in the same macroscopic volume, leading to an average `free' volume ${\langle V_{\rm f} \rangle}$; see Eqs.~(\ref{eq:ftrans_MCP})-(\ref{eq:VfMCP}). 
This correlation between the different ion species breaks the linear mixing rule.
At lower temperatures (see panel (a) of Fig.~\ref{fig:ppMCP_focp_mcpvsocp_with_modified_ftrans}), the deviation between the MCP and OCP predictions is negligible at densities below about 0.05~fm$^{-3}$, implying that the influence of the non-linear mixing term is not very important.
On the other hand, from panels (b) and (c) of Fig.~\ref{fig:ppMCP_focp_mcpvsocp_with_modified_ftrans} we can see that, with increasing temperature, the discrepancy between the two approaches starts to become significant at progressively lower densities.
This can be explained by the fact that the contribution of the translational free energy becomes more important at higher $T$ and $n_B$, as discussed in \citet{Dinh2022}.
In particular, we can see that the distributions in the (perturbative) MCP are shifted to larger clusters with respect to the OCP solution, and that the crust--core transition occurs earlier. 
This is because Eq.~(\ref{eqocp:rn}), which results in small clusters in the OCP approximation (see \citet{Dinh2022}), no longer holds in the MCP, as discussed in Sect.~\ref{sec:mostprobable}. 
In addition, the contribution of the translational free-energy term in the OCP approximation is more important than that in the MCP. 
Indeed, in the former, the free volume term is a variable entering the minimisation procedure, whereas $\langle V_{\rm f} \rangle$ in the MCP is the common volume for all nuclear species. As a result, it does not affect the nuclear distribution; see Eqs.~(\ref{eq:omegaj})--(\ref{eqMCP:Fij0star}).

\begin{figure}
    \centering \includegraphics[scale = 0.6]{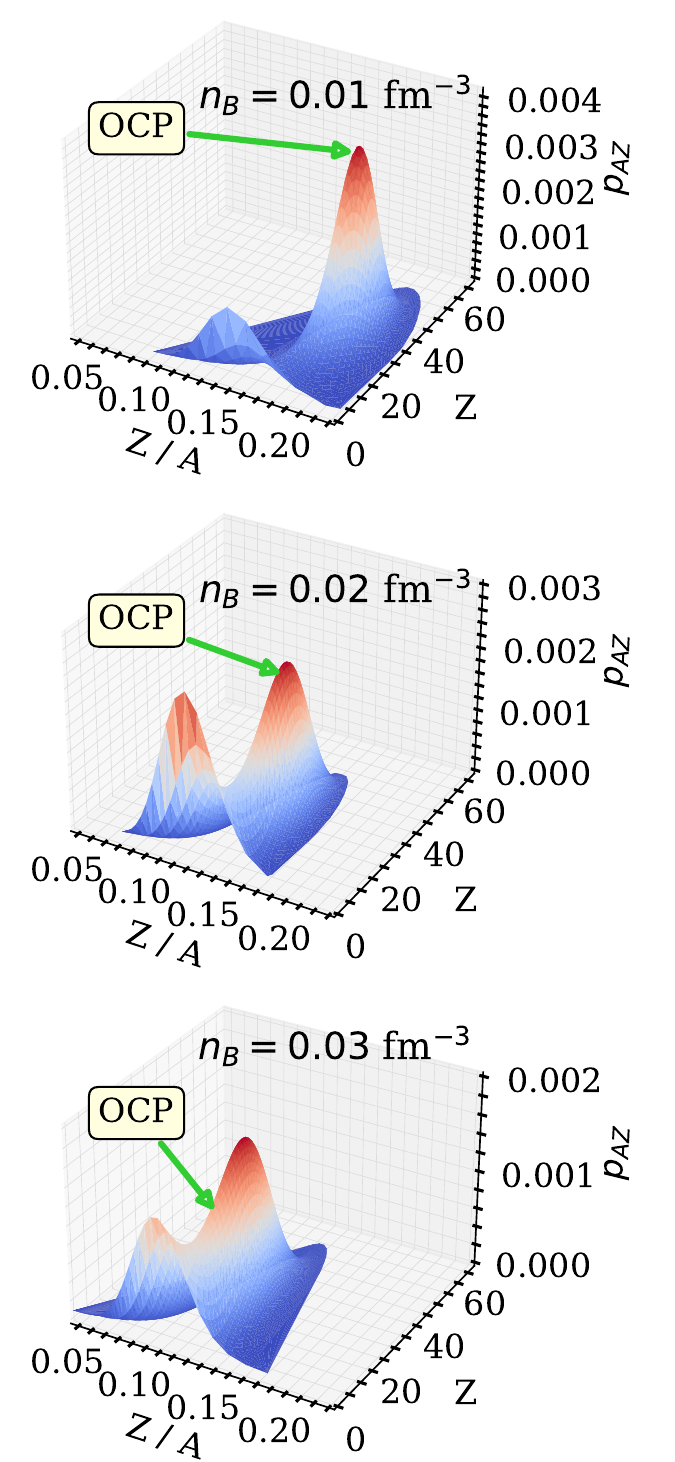}
    \caption{Normalised probability  $p_{AZ}$ in a perturbative MCP calculation as a function of the cluster proton number $Z$ and cluster proton fraction $Z/A$. Results were obtained at the selected temperature of $k_{\rm B}T = 2.0$~MeV for three different baryon densities: $n_B = 0.01$~fm$^{-3}$ (top panel), $n_B = 0.02$~fm$^{-3}$ (middle panel), and $n_B = 0.03$~fm$^{-3}$ (bottom panel). The green arrow in each panel indicates the corresponding OCP solution. 
    The translational free energy is included in both MCP and OCP calculations. 
   }
    \label{fig:pertubative_MCP_distributions_2MeV}
\end{figure}

\begin{figure}
    \centering
    \includegraphics[scale = 0.4]{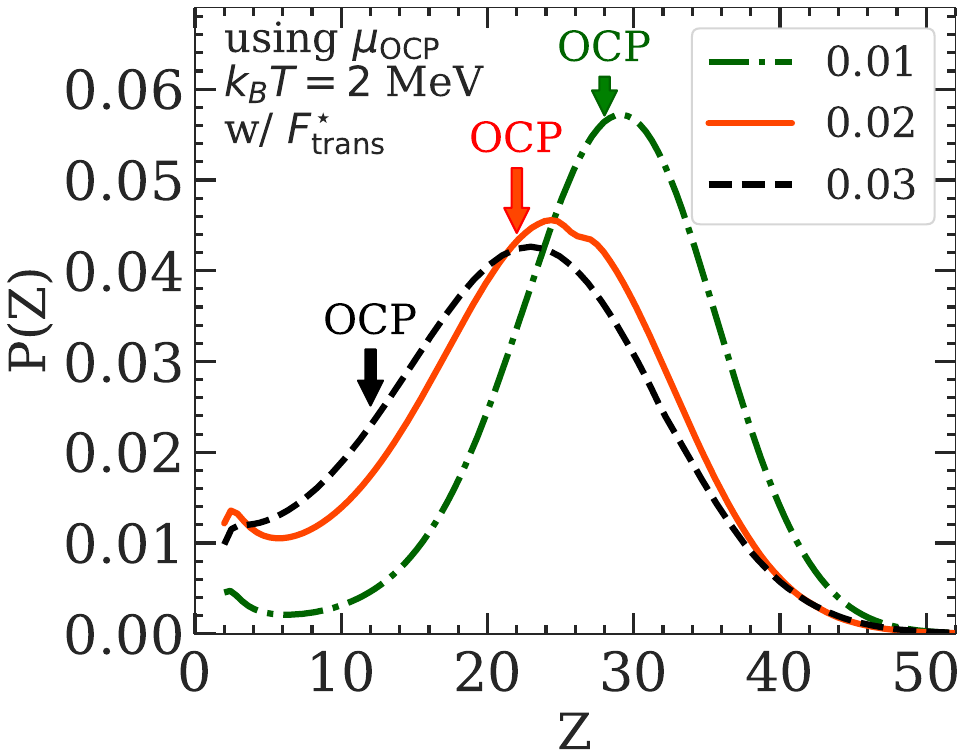}
    \caption{Normalised distributions $P(Z) = \sum_{A}p_{AZ}$, where $p_{AZ}$ are taken from Fig.~\ref{fig:pertubative_MCP_distributions_2MeV}, as a function of $Z$ in a perturbative MCP calculation. Calculations are performed for $k_{\rm B} T = 2$~MeV and $n_B = 0.01$~fm$^{-3}$ (green dash-dotted line), $n_B = 0.02$~fm$^{-3}$ (red solid line), and $n_B = 0.03$~fm$^{-3}$ (black dashed line). Arrows indicate the OCP solutions.}
    \label{fig:PZ_pertubative_2MeV}
\end{figure}

The discontinuous behaviour of the most probable $(A,Z)$ cluster in Fig.~\ref{fig:ppMCP_focp_mcpvsocp_with_modified_ftrans} is due to the fact that, as the temperature increases, the contribution of very light clusters, such as neutron-rich helium isotopes, becomes increasingly favoured. 
As shown in Fig.~\ref{fig:pertubative_MCP_distributions_2MeV}, the distributions become double-peaked when the temperature overcomes $k_{\rm B} T \approx 1$ MeV, and a density domain exists where the helium abundance overcomes that of the heavy cluster in the iron region predicted by the OCP.
Still, it is also visible from Fig.~\ref{fig:pertubative_MCP_distributions_2MeV} that the light-cluster contribution is limited to a small number of helium isotopes, while a large variety of nuclear species have a comparable probability around the $(A^{\rm OCP},Z^{\rm OCP})$ value. 
Therefore, if we consider the probabilities of each element $Z$ by integrating the distributions over the isotopic content (i.e. over $A$), the (perturbative) MCP predictions appear to be in closer agreement with the OCP approximation, at least at relatively low densities, as can be seen from Fig.~\ref{fig:PZ_pertubative_2MeV}.

The results presented in this section confirm that the non-linear mixing term induced by the translational energy leads to a breakdown of the ensemble equivalence between the OCP and MCP predictions. 

The perturbative MCP approach employed here has the clear advantage of computing the full nuclear distributions at a reduced computational cost.
However, this approach is not fully self-consistent, because the gas variables and chemical potentials resulting from the OCP solutions might not exactly satisfy the constraints of baryon number conservation and charge neutrality in the MCP, Eqs.~(\ref{eqMCP:baryon}) and (\ref{eqMCP:charge}), respectively. 
For this reason, we computed the full self-consistent MCP, which is discussed in the following section.

\subsection{Self-consistent MCP calculations}
\label{sec:exact_MCP}

In this section, we present the results obtained by performing fully self-consistent MCP calculations.
At each thermodynamic condition defined by ($n_B$, $T$), we solved the system of equations given by Eqs.~(\ref{eqMCP:baryon}), (\ref{eqMCP:charge}), (\ref{eqMCP1:u}), (\ref{eqMCP1:mun_MCP}), and (\ref{eqMCP1:PMCP}), together with the equation for the ion density $n_N^\j$, Eq.~(\ref{eqMCP:nN}).
The translational free energy is always included in the computation of the ion free energy.

\begin{figure}
    \centering
    \includegraphics[scale = 0.4]{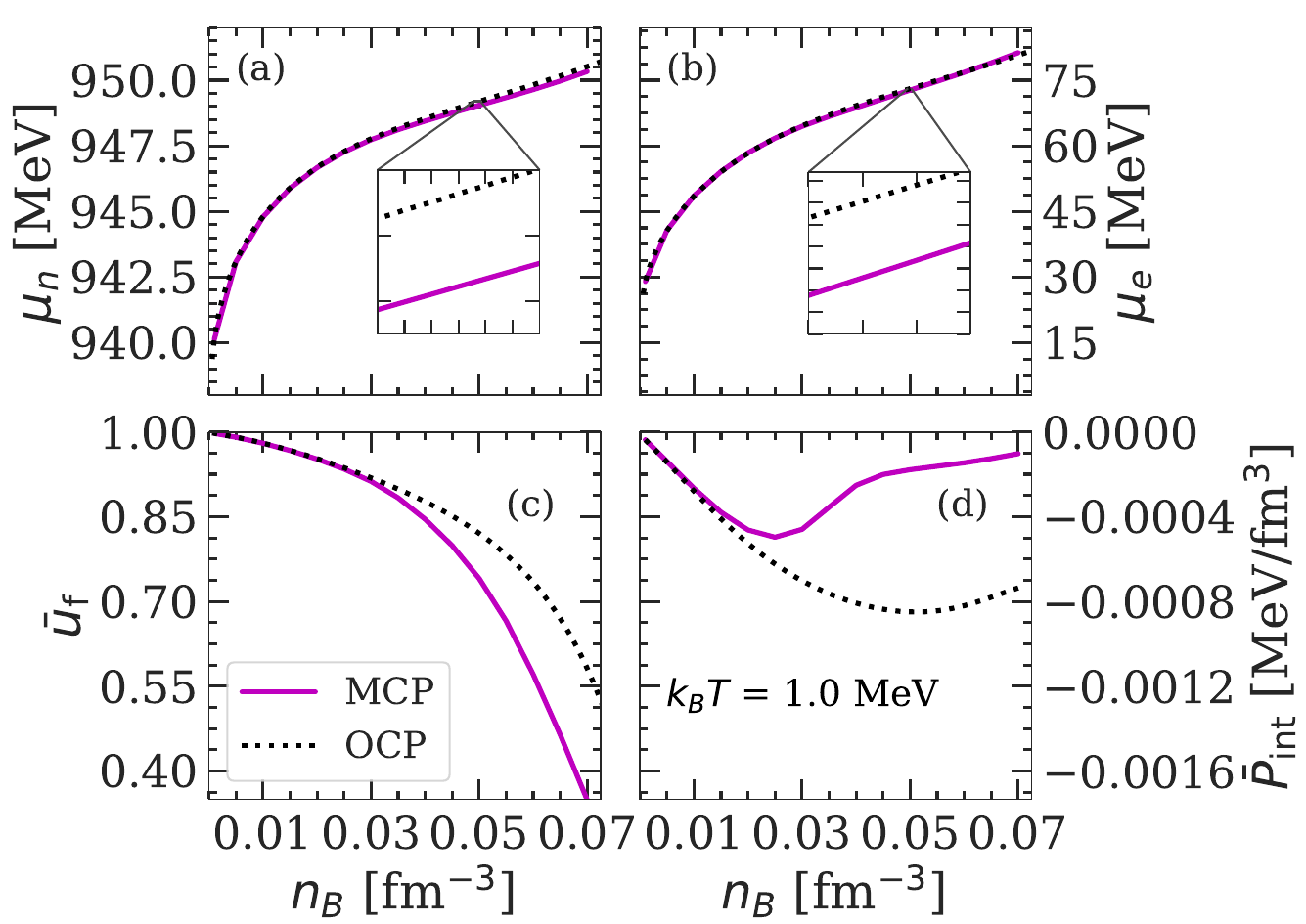}
    \caption{Self-consistent MCP solution (solid lines) for the neutron chemical potential ${\mu_{n}}$  (panel a), electron chemical potential $\mu_e$ (panel b), average free volume fraction $\Bar{u}_{\rm f}$ (panel c), and MCP interaction pressure $\Bar{P}_{\rm int }$ (panel d) as a function of the total baryon density $n_B$ at $k_{\rm B} T = 1$~MeV. The dotted lines show the OCP predictions. The tick marks of the x and y axes in the insets of panels (a) and (b) are spaced by $2\times10^{-4}$ fm$^{-3}$ and 0.1~MeV, respectively.
    }
    \label{fig:PPMCPspheres_MCPsolution}
\end{figure}

With Fig.~\ref{fig:PPMCPspheres_MCPsolution}, we begin  by showing the evolution with the baryon number density of the neutron chemical potential $\mu_n$ presented in Eq.~(\ref{eq:mun_MCP}) (panel a), the electron chemical potential potential $\mu_e = \mathrm{d} \mathcal{F}_e / \mathrm{d} n_e$ (panel b), the average free-volume fraction ${\Bar{u}_{\rm f}}$ in  Eq.~(\ref{eqMCP1:u}) (panel c), and the MCP interaction pressure $\Bar{P}_{\rm int}$ presented in Eq.~(\ref{eqMCP1:PMCP}) (panel d).
All these quantities enter the computation of the ion abundances via Eq.~(\ref{eqMCP:tidle_omegaij}).
For illustrative purposes, we only plot the results at one selected temperature, $k_{\rm B}T = 1$~MeV. 
The solutions from the MCP calculations are shown by solid lines, while the dotted lines correspond to the values calculated at the OCP composition, that is, $\mu_n^{\rm OCP}, \mu_e^{\rm OCP}, \Bar{u}_{\rm f}^{\rm OCP}$, and $\Bar{P}_{\rm int}^{\rm OCP}$.
As we can see from panels (a)\ and (b), the chemical potentials in the MCP are very similar to those in the OCP approximation; indeed, the two curves are almost indistinguishable.
However, from the inset in panel (a) we can notice that the neutron chemical potential in the MCP is actually slightly lower than the OCP counterpart, suggesting that the corresponding gas density is not the same in the two approaches, and in particular that it is smaller in the MCP. 
The same trend is observed for the electron chemical potential (see panel (b) in Fig.~\ref{fig:PPMCPspheres_MCPsolution}), except at densities above $~ 0.06$~fm$^{-3}$, where $\mu_e$ calculated in the MCP is slightly above that obtained in the OCP approach, and therefore the corresponding density is higher in the MCP.
The lower density of the unbound nucleons in the MCP was also observed by \citet{Burrows1984} and by \citet{gulrad2015} (see their Fig.~13).
Although the difference between the MCP and OCP chemical potentials is numerically small, given that these latter enter in the calculation of the ion abundances through the exponential, Eq.~(\ref{eqMCP:nN}), this difference could still lead to a significant deviation between the MCP and OCP results, as already noted by \citet{gulrad2015}. 
From panels (c) and (d) in Fig.~\ref{fig:PPMCPspheres_MCPsolution}, one can see that at relatively low baryon density, $n_B \lesssim 0.02$ fm$^{-3}$,
${\Bar{u}_{\rm f}}$ and $\Bar{P}_{\rm int}$ computed within the MCP are almost identical to those calculated in the OCP approximation\footnote{We notice that  ${\Bar{u}^{\rm OCP}_{\rm f}} = u_{\rm f} \equiv V_{\rm f}/V_{\rm WS} \neq V^{\rm OCP}_{\rm f}/V_{\rm WS}$ due to the different definition of $V_{\rm f}$; see Eq.~(\ref{eq:VfMCP}) and Eq.~(\ref{eq:Vf}). 
}. 
We therefore expect that, at low density and temperature, the results shown in Fig.~\ref{fig:ppMCP_focp_mcpvsocp_with_modified_ftrans} are a good approximation. 
On the other hand, at higher densities, the absolute value of the interaction pressure, which directly enters into the computation of the rearrangement term, tends to zero in the MCP approach, while $|\Bar{P}_{\rm int}^{\rm OCP}|$ increases. 
As the latter is an input in the perturbative MCP, we might also expect the discrepancies between the perturbative and self-consistent MCP approaches to become more pronounced at higher densities. 

\begin{figure}
    \centering
    \includegraphics[scale = 0.4]{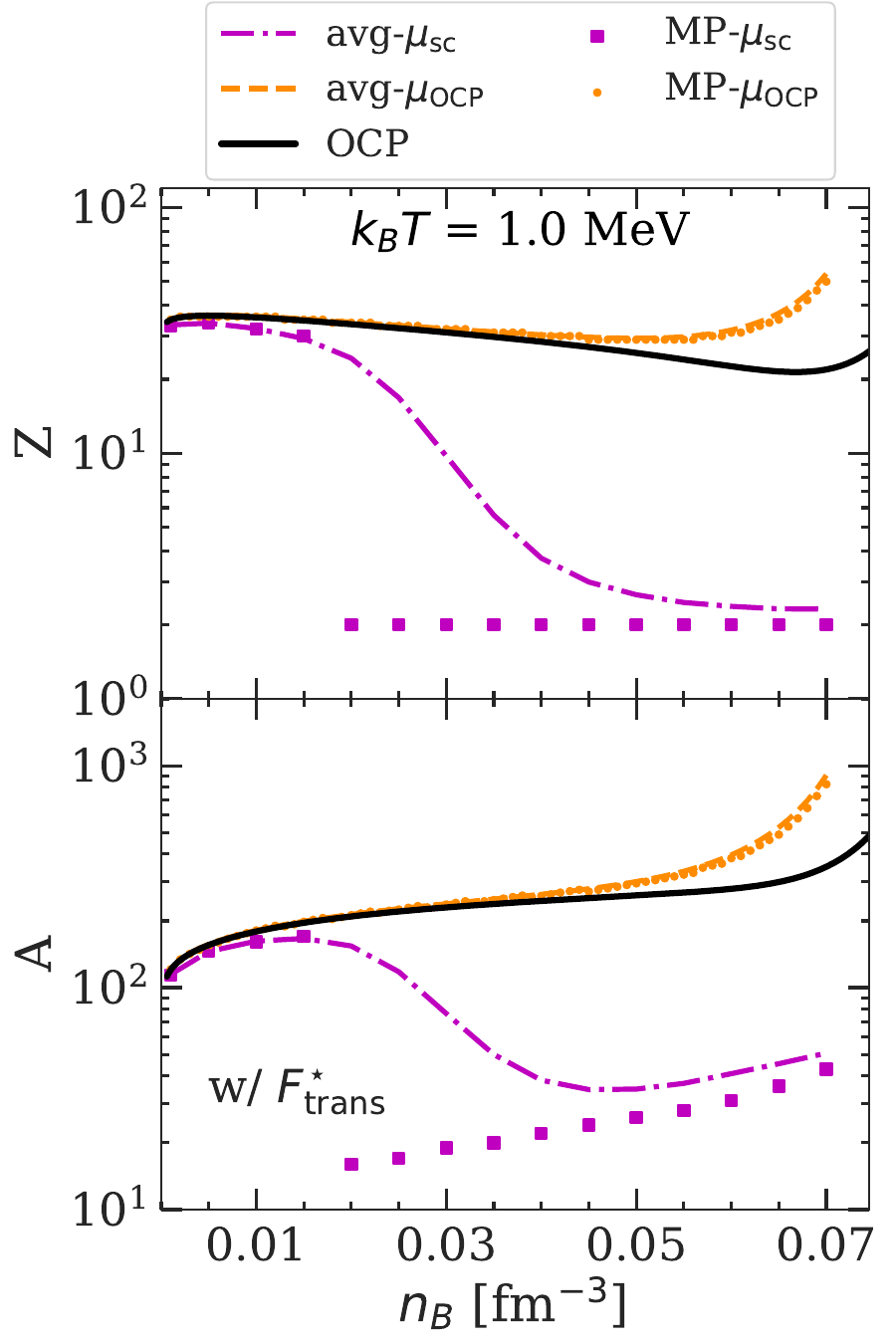}
    \caption{Average (dash-dotted violet lines, labelled `avg-$\mu_{\rm sc}$') and most probable (violet squares, labelled `MP-$\mu_{\rm sc}$') values of the cluster proton number $Z$ (top panel) and mass number $A$ (lower panel) in the self-consistent MCP calculations as a function of the total baryon density $n_B$ at $k_{\rm B} T = 1$~MeV. For comparison, the OCP solutions (black solid lines), as well as the average (orange dashed lines labelled `avg-$\mu_{\rm OCP}$') and most probable (orange dots labelled `MP-$\mu_{\rm OCP}$') values in the perturbative MCP are also plotted.}  
    \label{fig:ppMCP_NSE_mcpvsocp_with_modified_ftrans}
\end{figure}

Indeed, this is shown in Fig.~\ref{fig:ppMCP_NSE_mcpvsocp_with_modified_ftrans}, where the average (dash-dotted violet lines) and the most probable (violet squares) values of $Z$ (top panel) and $A$ (lower panel) in the self-consistent MCP calculations are plotted as a function of $n_B$ at $k_{\rm B} T = 1$~MeV, together with the OCP predictions (black solid lines) and the average (orange dashed lines) and most probable (orange dots) values obtained within the perturbative MCP procedure. 
Below about $0.02$~fm$^{-3}$, the results of the three different approaches almost coincide. 
However, as the density increases, the outcomes of the different methods start to diverge, and therefore the perturbative MCP and OCP predictions become less reliable. 
 On the one hand, these results confirm the validity of these approximations at relatively low densities, while on the other, they highlight the importance of full MCP calculations in the deeper region of the PNS inner crust.

\begin{figure*}
    \centering
    \includegraphics[scale = 0.4]{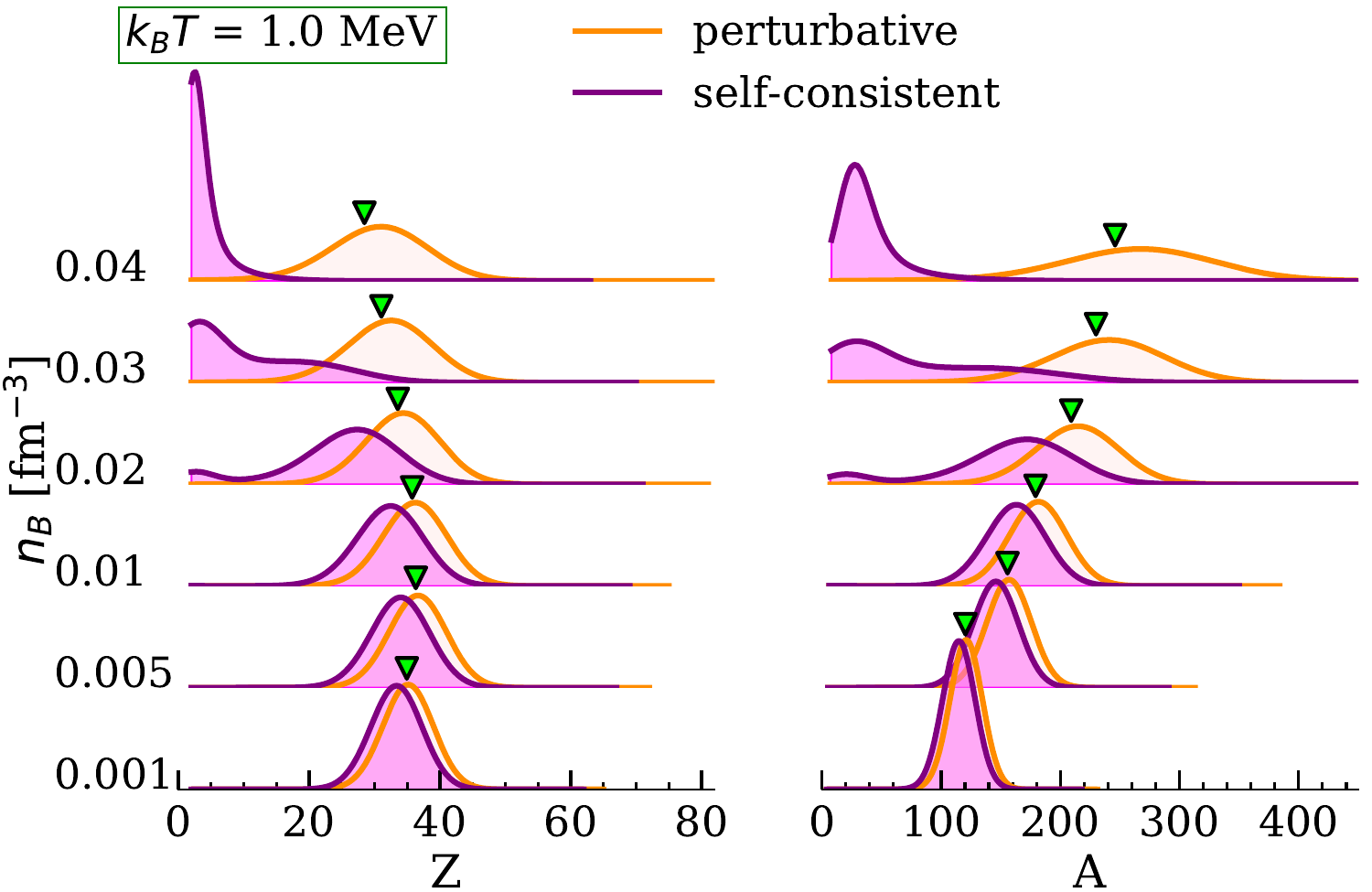}
    \caption{Normalised probability distributions of the cluster proton number $Z$ (left panel), $\sum_{A}p_{AZ}$, and mass number $A$ (right panel), $\sum_{Z}p_{AZ}$, at $k_{\rm B} T = 1$~MeV for different baryon densities, $n_B \in [0.001, 0.04]$~fm$^{-3}$. The violet distributions are obtained from the self-consistent MCP calculations, while the orange ones correspond to the perturbative MCP. The OCP solutions are marked by the green triangles.}
    \label{fig:ppMCPspheres_NSE_distribution_vs_nb}
\end{figure*}
\begin{figure*}
    \centering
    \includegraphics[scale = 0.4]{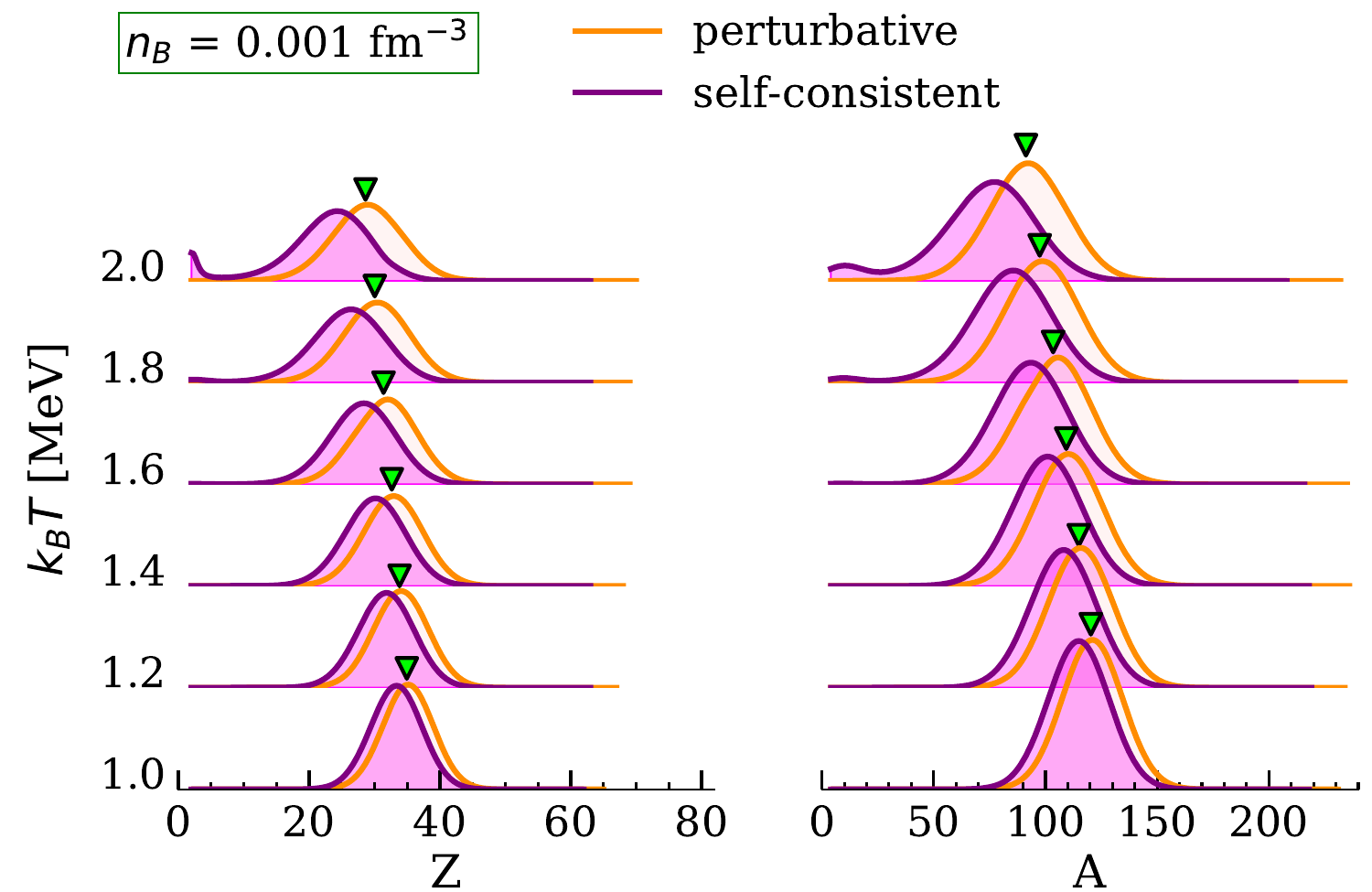}
    \caption{Same as Fig.~\ref{fig:ppMCPspheres_NSE_distribution_vs_nb}, but at a fixed baryon density $n_B = 0.001$~fm$^{-3}$ and for different temperatures, $k_{\rm B} T \in [1.0, 2.0]$ MeV.}
    \label{fig:ppMCPspheres_NSE_distribution_vs_T}
\end{figure*}

From Fig.~\ref{fig:ppMCP_NSE_mcpvsocp_with_modified_ftrans}, it is also interesting to see that, in the self-consistent MCP calculations, lighter nuclei dominate at high density, which is in agreement with previous works based on the nuclear statistical equilibrium (see e.g. \citet{Souza2009,gulrad2015}). 
Conversely, in the OCP and perturbative MCP approximations, heavier clusters survive until the crust--core transition.

In order to further investigate this point, we plot in Fig.~\ref{fig:ppMCPspheres_NSE_distribution_vs_nb} the normalised distributions of the cluster proton number $Z$ (left panel), $\sum_{A}p_{AZ}$, and mass number $A$ (right panel), $\sum_{Z}p_{AZ}$, at different densities in the inner crust, $n_B \in [0.001, 0.04]$~fm$^{-3}$, at $k_{\rm B} T = 1$~MeV, for both the self-consistent MCP calculations (violet areas) and the perturbative MCP ones (orange areas). 
As one may expect, at lower densities, the distributions obtained in the two approaches are almost identical; they are narrow and peaked at the OCP predictions (green triangles). 
Therefore, employing a perturbative MCP where the gas variables are fixed from the OCP solution is a very good approximation in these thermodynamic conditions.
As the density increases, at $n_B \approx 0.02 - 0.03$~fm$^{-3}$,
the self-consistent MCP distributions are displaced towards lower values of $A$ and $Z$, and even exhibit a double-peak structure, meaning that lighter clusters coexist with heavier ones and their contribution becomes important.
In these regimes, the OCP and perturbative MCP predictions significantly overestimate the cluster mass and proton number. 
Eventually, at the bottom of the crust, nuclei with $Z < 5$ and $A<50$ may even dominate, with helium being the most probable element, although the tails of the distributions extend up to $Z \simeq 50$ and $A \simeq 400$, implying that heavier nuclei may still be present near the crust--core transition. 
However, it is to be noted that, while the $Z=2$ element appears to be the most abundant, $^4$He, which is the most bound helium isotope in the vacuum, is not the most abundant cluster (see also Fig.~\ref{fig:ppMCP_NSE_mcpvsocp_with_modified_ftrans}).
This is because in the NS inner crust, the medium is very dense and (helium) clusters are immersed in continuum neutron states.
Indeed, as already noted by \citet{gulrad2015}, alpha particles are abundant for matter close to isospin symmetry, while more neutron-rich (hydrogen and helium) isotopes prevail in neutron-rich matter.

The findings presented above were obtained at a given temperature, namely $k_{\rm B} T = 1$~MeV. 
In order to study the dependence of the results on the temperature, we ran the calculations for different temperatures up to $k_{\rm B} T = 2$~MeV.
In Fig.~\ref{fig:ppMCPspheres_NSE_distribution_vs_T}, we plot the distributions of $Z$ (left panel) and $A$ (right panel), defined as in Fig.~\ref{fig:ppMCPspheres_NSE_distribution_vs_nb}, at $n_B = 0.001$~fm$^{-3}$ in the aforementioned temperature range.
As in Fig.~\ref{fig:ppMCPspheres_NSE_distribution_vs_nb}, the results from both the self-consistent  MCP calculations (violet areas) and the perturbative MCP ones (orange areas) are displayed and compared with the OCP solutions (green triangles).
At lower temperatures, the peaks of the distribution coincide with or are very close to the OCP predictions. 
However, as the temperature increases, more nuclear species are populated, and therefore the distributions flatten and the OCP prediction tends to overestimate the cluster size. 
At higher temperature, $k_{\rm B} T \geq 1.8$~MeV, small clusters start to appear.
At 2~MeV, the double-peak structure already observed at high density is clearly visible in the self-consistent MCP calculations, while it is absent in the perturbative MCP results.
Nevertheless, the deviation between the predictions of the two approaches is not as significant as in Fig.~\ref{fig:ppMCPspheres_NSE_distribution_vs_nb}, which suggests that the effect of density overcomes that of temperature as far as the cluster distributions are concerned.

\begin{figure}
    \centering
    \includegraphics[scale = 0.4]{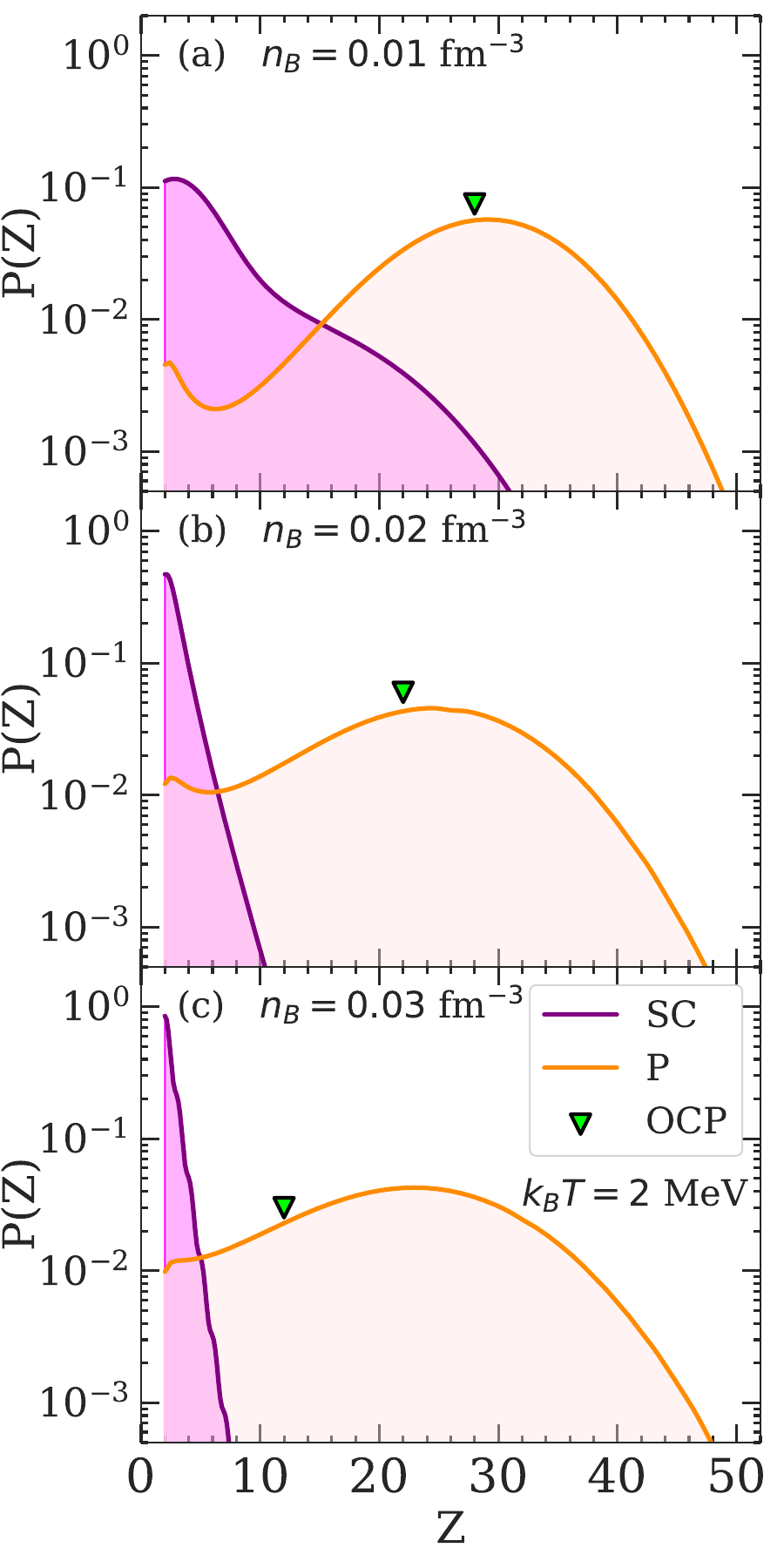}
    \caption{Normalised distributions $P(Z) = \sum_{A}p_{AZ}$ as a function of $Z$ in a fully self-consistent (`SC') and a perturbative (`P') MCP calculation. Calculations are performed for $k_{\rm B} T = 2$~MeV and $n_B = 0.01$~fm$^{-3}$ (panel a), $n_B = 0.02$~fm$^{-3}$ (panel b), and $n_B = 0.03$~fm$^{-3}$ (panel c). Green triangles indicate the OCP solutions.}
    \label{fig:PZ_pert-sc}
\end{figure}

A complementary comparison of the distributions obtained in the perturbative and self-consistent MCP is given in Fig.~\ref{fig:PZ_pert-sc}, where the normalised distributions $\sum_A p_{AZ}$ are plotted as a function of $Z$ for $k_{\rm B}T = 2$~MeV and for three selected densities: $n_B = 0.01$~fm$^{-3}$ (panel a), $n_B = 0.02$~fm$^{-3}$ (panel b), and $n_B = 0.03$~fm$^{-3}$ (panel c).
These conditions correspond to those for which a double-peaked distribution is observed in the perturbative MCP calculations (see Fig.~\ref{fig:pertubative_MCP_distributions_2MeV}).
For comparison, the OCP results are also shown (green triangles).
At all three densities considered, the distributions obtained with the self-consistent MCP are shifted towards lower $Z$ with respect to those of the perturbative MCP calculations, as already observed in Fig.~\ref{fig:ppMCPspheres_NSE_distribution_vs_nb} at $k_{\rm B}T = 1$~MeV.
Indeed, at $k_{\rm B}T = 2$~MeV, in the self-consistent MCP, light nuclei are always more abundant than heavier ones.
Moreover, the distributions become peaked around $Z \approx 2$ (see panels (b) and (c)), while they remain relatively broad and peaked towards $Z \approx 30$ in the perturbative MCP.

\begin{figure}
    \centering
    \includegraphics[scale = 0.4]{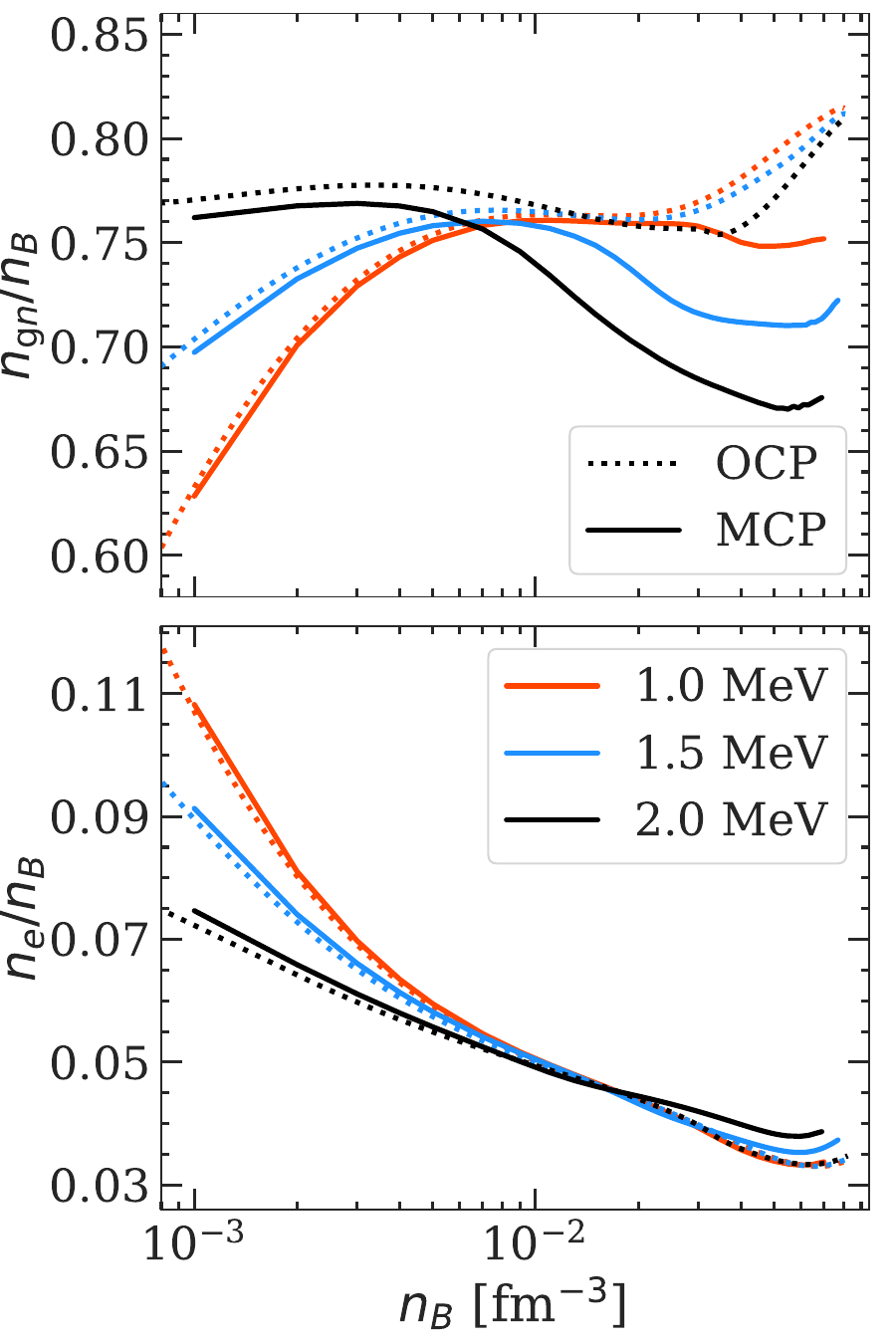}
    \caption{Ratio of the neutron gas density (upper panel) and of the electron gas density (lower panel) over the baryon density, $\nng/n_B$ and $n_e/n_B$, respectively, as a function of the total baryon density $n_B$ resulting from the self-consistent MCP calculations (solid lines) at $k_{\rm B} T = 1$~MeV (red lines), $1.5$~MeV (blue lines), and $2$~MeV (black lines). For comparison, the OCP results are plotted with dotted lines.}
    \label{fig:gas}
\end{figure}

\begin{figure}
    \centering
    \includegraphics[scale = 0.4]{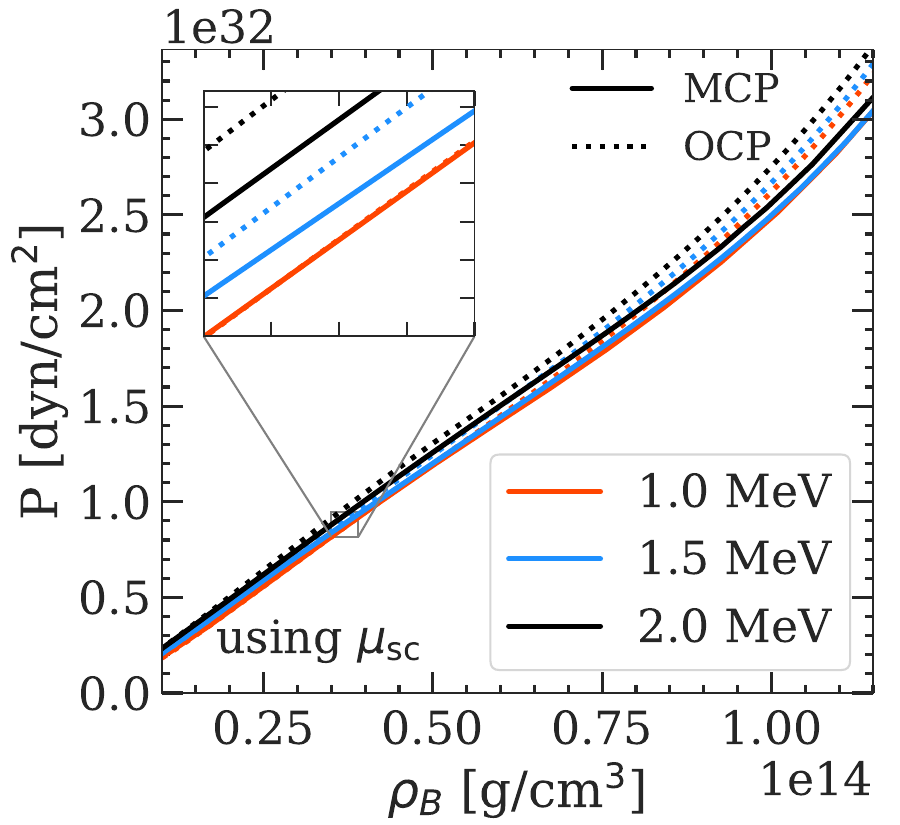}
    \caption{Total pressure $P$ vs. mass-energy density $\rho_B$ resulting from the self-consistent MCP calculations (solid lines) at $k_{\rm B} T = 1$~MeV (red lines), $1.5$~MeV (blue lines), and $2$~MeV (black lines). For comparison, the OCP results are plotted with dotted lines. 
    In the inset, the tick marks on the x and y axis are spaced by $10^{12}$ g/cm$^3$ and $2\times 10^{30}$ dyn/cm$^2$, respectively.}
    \label{fig:EoS}
\end{figure}

The results presented in Figs.~\ref{fig:ppMCP_NSE_mcpvsocp_with_modified_ftrans}-\ref{fig:PZ_pert-sc} show that even for temperatures as low as 1~MeV (and more so at higher temperatures), the OCP and the perturbative MCP approximations are no longer reliable in the deepest region of the crust. 
Therefore, for studies requiring accurate knowledge of the crust composition, a full MCP calculation is needed.
However, the effect is less important as far as more global properties such as the equation of state are concerned.
This is illustrated in Fig.~\ref{fig:gas}, where we display the neutron and electron gas density over the baryon number density, $\nng/n_B$ and $n_e/n_B$, respectively, and in Fig.~\ref{fig:EoS}, where we plot the total pressure versus the mass-energy density in the PNS inner crust.
Results are shown for three different temperatures, $k_{\rm B} T = 1$~MeV (red lines), $1.5$~MeV (blue lines), and $2$~MeV (black lines), for both the self-consistent MCP (solid lines) and the OCP approximation (dotted lines).
From Fig.~\ref{fig:gas}, we can see that the MCP neutron gas densities are slightly smaller than the OCP ones in all the considered cases, although differences of up to about $20\%$ are observed for $k_{\rm B} T = 2$~MeV at the highest densities in the crust.
As for the electron density, that calculated within the MCP is generally slightly higher than that calculated in the OCP approximation, except possibly at $k_{\rm B} T = 1$~MeV around a few $10^{-2}$~fm$^{-3}$, in accordance with Fig.~\ref{fig:PPMCPspheres_MCPsolution}.
From Fig.~\ref{fig:EoS}, we note that, concerning the equation of state, at $k_{\rm B} T = 1$~MeV, the pressure-density relations provided within the self-consistent MCP approach and the OCP approximation are very similar at all densities, except very close to the crust--core transition. 
At higher temperatures, deviations between the two calculations start to emerge, particularly at high density.
However, these discrepancies amount to approximately $10\%$ at most in the vicinity of the crust--core transition.

\subsection{Impurity parameter}
\label{sec:Qimp}

\begin{figure}
    \centering
    \includegraphics[scale = 0.4]{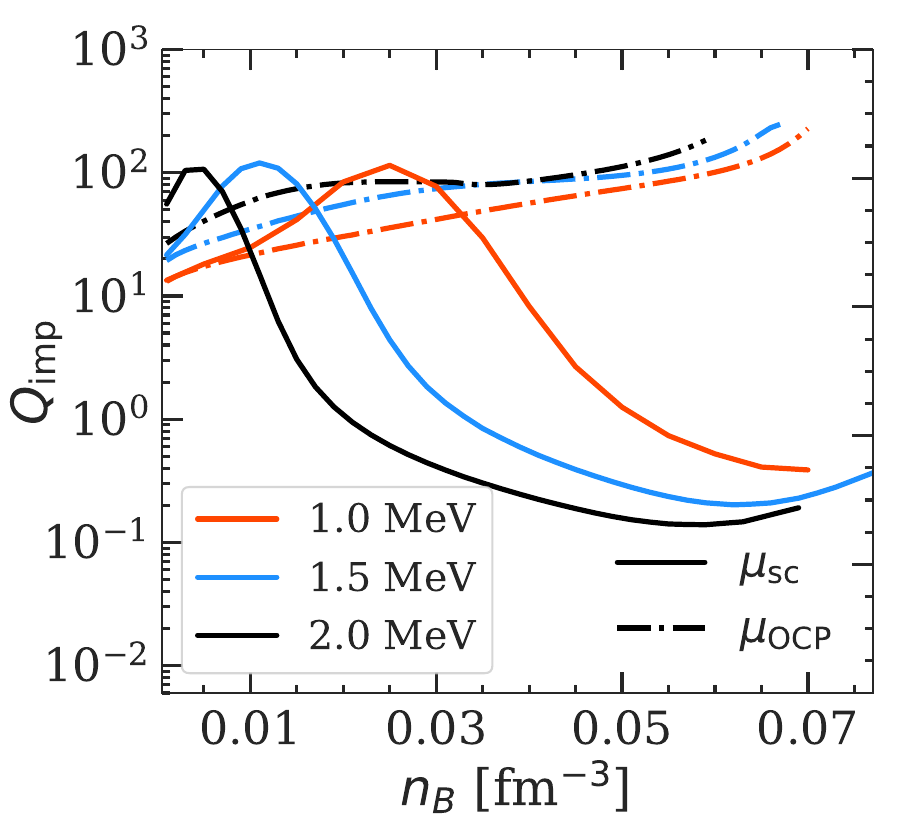}
    \caption{Impurity parameter $Q_{\rm imp}$ as a function of the baryon density $n_B$ in the inner crust for $k_{\rm B}T = 1.0$ MeV (red lines), $k_{\rm B}T = 1.5$ MeV (blue lines), and $k_{\rm B}T = 2.0$ MeV (black lines) in the full MCP calculation (solid lines). For comparison, results obtained with the perturbative MCP (dash-dotted lines) are also shown.} 
    \label{fig:Qimp}
\end{figure}

Within our MCP approach, it is also possible to self-consistently calculate the impurity parameter, which is defined as the variance of the $Z$ distribution,
\begin{equation}
    Q_{\rm imp} = \langle Z^2 \rangle - \langle Z \rangle^2 \ . 
\end{equation}
The latter is plotted in Fig.~\ref{fig:Qimp} for three temperatures, $k_{\rm B}T = 1.0$~MeV (red lines), $k_{\rm B}T = 1.5$~MeV (blue lines), and $k_{\rm B}T = 2.0$~MeV (black lines).
Results obtained with the self-consistent (perturbative) MCP calculations as a function of the baryon density $n_B$ in the crust are shown as solid (dash-dotted) lines. 
At the lower temperature of $k_{\rm B}T = 1.0$~MeV (red lines), the predictions from the two treatments coincide at lower densities, until $n_B \approx 0.01$~fm$^{-3}$. 
With increasing density, the impurity parameter is, at first, larger in the self-consistent MCP approach with respect to the perturbative one.
However, at variance with the perturbative MCP predictions, $Q_{\rm imp}$ does not increase monotonically in the self-consistent calculations, but peaks around $n_B \approx 0.025$~fm$^{-3}$, reaching $Q_{\rm imp}\approx 100$ for the considered BSk24 model, and subsequently decreases at higher densities.
A similar behaviour is observed for higher temperatures, although the discrepancy between the two treatments and the peak in the impurity parameter appear at smaller densities.
This can be understood from the $Z$ distributions shown in the left panel of Figs.~\ref{fig:ppMCPspheres_NSE_distribution_vs_nb}-\ref{fig:ppMCPspheres_NSE_distribution_vs_T} and in Fig.~\ref{fig:PZ_pert-sc}.
Indeed, while the presence of the second peak at small $Z$ for moderate densities increases the variance of $Z$, the transition to light nuclei at high density overall decreases the value of the impurity parameter.
These findings also show that the impurity parameter calculated within the perturbative MCP approach is underestimated at low densities, and is severely overestimated at high densities.

\begin{figure}
    \centering
    \includegraphics[scale  = 0.4 ]{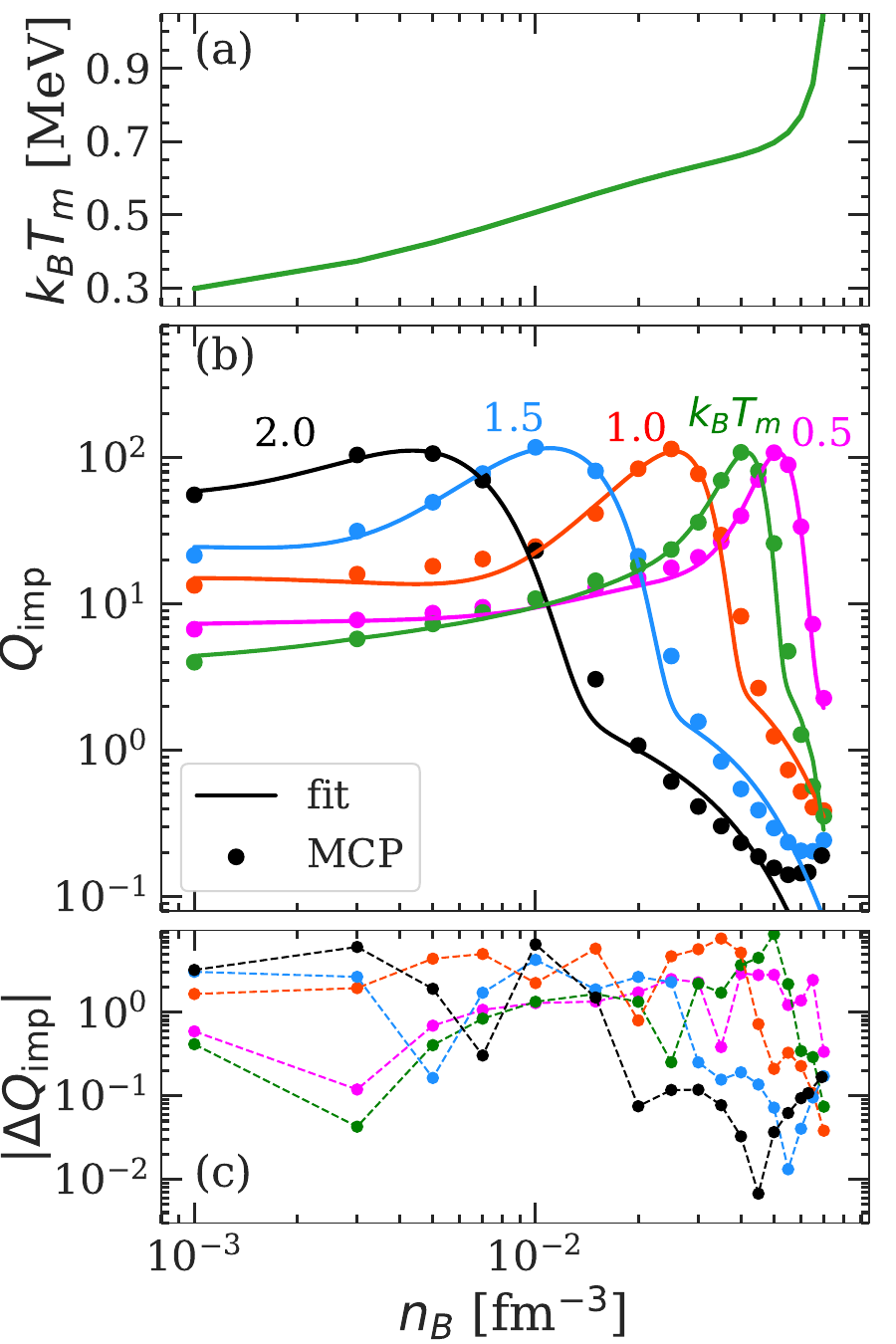}
    \caption{Impurity parameter at different temperatures. Panel (a): Crystallisation temperature $k_{\rm B} T_{\rm m}$ calculated using Eq.~(\ref{eq:Tm}) as a function of the baryon density $n_B$.  
    Panel (b): Impurity parameter $Q_{\rm imp}$ in the fully self-consistent MCP calculation as a function of $n_B$ evaluated at the crystallisation temperature $k_{\rm B} T_{\rm m}$ (green points) and at four fixed temperatures: $k_{\rm B} T = 0.5$~MeV (magenta points), $k_{\rm B} T = 1.0$~MeV (red points), $k_{\rm B} T = 1.5$~MeV (blue points), and $k_{\rm B} T = 2.0$~MeV (black points). Solid lines are obtained from the fit given by Eq.~(\ref{eq:fit_Qimp}). 
    Panel (c): Absolute error in the fitting formula ($|\Delta Q_{\rm imp}|$ = |calculated $Q_{\rm imp}$ - fit|).}
    \label{fig:Qimp_Tm}
\end{figure}

In the cooling process of a NS, it is reasonable to assume that the crust composition is frozen after the solidification of the crust. 
Neutron absorption or $\beta$-decays might still occur at lower temperatures, but pycnonuclear reactions that involve overcoming a Coulomb barrier might be considerably inhibited even above crystallisation. A realistic estimate of the temperature at which the ion distribution is frozen and the impurity parameter is settled would require a comparison between the cooling time and the different reaction rates, which is beyond the scope of the present study. 
As a reasonable estimate, the impurity parameter of the solid crust can be computed from that obtained at the crystallisation temperature of a pure Coulomb plasma \citep{hpy2007}:
\begin{equation}
    T_{\rm m} \approx \frac{Z^2e^2}{k_{\rm B} r_{\rm WS}\Gamma_{\rm m}},
    \label{eq:Tm}
\end{equation}
where $Z$ and $r_{\rm WS}$ correspond to the ground-state composition at zero temperature, and $\Gamma_{\rm m} \approx 175$ is the Coulomb coupling parameter at the melting point.
It is interesting to observe that this simple expression was shown in \citet{Carreau2020a} (see their Fig.~4) to give a good order-of-magnitude estimate of the temperature where the OCP free-energy densities in the liquid equate to those of the solid phase.

Values of $k_{\rm B} T_{\rm m}$ from Eq.~(\ref{eq:Tm}) are displayed in panel (a) of Fig.~\ref{fig:Qimp_Tm}: $k_{\rm B} T_{\rm m}$ increases monotonically from $\sim 0.3$~MeV at $n_B = 10^{-3}$~fm$^{-3}$ to $\sim 1.0$~MeV at $n_B = 0.07$~fm$^{-3}$ (see also Fig.~4 in \citet{Carreau2020a}).
The corresponding evolution of the impurity parameter in the fully self-consistent MCP calculations at the crystallisation point as a function of the baryon density in the inner crust is shown with green points in panel (b) of Fig.~\ref{fig:Qimp_Tm}. 
We can see that a high impurity parameter $10 \lesssim Q_{\rm imp}\lesssim 100$ should be expected in the whole inner crust, which could have important consequences for the magneto-thermal evolution of X-ray pulsars (see \citet{Pons2013,Newton2013b,Horowitz2015}).

For comparison, we also plot $Q_{\rm imp}$ for four different selected temperatures: $k_{\rm B} T = 0.5$~MeV (magenta points), $k_{\rm B} T = 1.0$~MeV (red points), $k_{\rm B} T  = 1.5$~MeV (blue points), and $k_{\rm B} T = 2.0$~MeV (black points). 
We can see that the general behaviour of $Q_{\rm imp}$ is similar for all temperatures; that is, all curves show a peak and a subsequent drop. 
Indeed, even at a relatively low temperature of $k_{\rm B} T = 0.5$~MeV, $Q_{\rm imp}$ is still reduced at high densities, $n_B \approx 0.06$~fm$^{-3}$, indicating that the charge distribution is dominated by small $Z$.
Moreover, as expected, a higher $Q_{\rm imp}$ is obtained if the composition is frozen at higher temperature, but the sharp drop occurs at lower density because of the appearance of light nuclei (see also Figs.~\ref{fig:ppMCPspheres_NSE_distribution_vs_nb}-\ref{fig:PZ_pert-sc}).
However, our results should be considered with care at high density, because non-spherical pasta structures could appear, and are not considered in the present paper. 
Such structures are also expected to be associated with high impurity parameter values $Q_{\rm imp}\sim 30$; see e.g. \citet{Schneider2016} and \citet{Caplan2021}.
Also, the CLD description can be questioned for light clusters and a more microscopic treatment of the in-medium modifications might be needed; see \citet{Pais2018} and \citet{Roepke2020}.

The results for the impurity parameters shown in Fig.~\ref{fig:Qimp_Tm} are given in Table~\ref{tab:Qimp} and are available in tabular format at the CDS.
For practical applications to numerical simulations, we also provide a fitting formula:
\begin{equation}
    Q_{\rm imp} = \exp \left(a_0 + \sum_{k=0}^4 x_B^k \sum_{j=1}^3 a_{kj} x_T^j \right) + \Delta Q \ ,
 \label{eq:fit_Qimp}
\end{equation}
where $x_B \equiv n_B /\text{fm}^{-3}$, $x_T \equiv k_{\rm B}T/\text{MeV}$, $a_0 = 3.328$, and $a_{kj}$ are given in Table~\ref{tab:Qimp_fit_params}, and the function $\Delta Q$ reads 
\begin{equation}
     \Delta Q =  \frac{f_0\ x_T\ (x_{BT}(n_B,T))^{f_1}} {\left[(x_{BT}(n_B, T) / f_2 + f_3)^4 + f_4 \right]f(n_B,T)},
     \label{eq:deltaQimp}
\end{equation}
with $f_0 = 65.55 $, $f_1 = 9.046 \times 10^{-2}$, $f_2 = -6.324$, $f_3 = -1.087 \times 10^{-2}$, and $f_4 = 4.802$. 
The variable $x_{BT}$ in Eq.~(\ref{eq:deltaQimp}) depends on both $n_B$ and $T$ and is defined as
\begin{equation}
    x_{BT}(n_B,T) \equiv \left( 5 + \log_{10} x_B \right)^2x_T \ ,
\end{equation}
while $f(n_B,T)$ is given by
\begin{equation}
 f(n_B,T) = 1 +  x_Bx_T^{3/2}  \exp \left(\frac{200}{3}x_B - \frac{1}{4} \right).
\end{equation}
The impurity parameter calculated at different temperatures with the fitting expression Eq.~(\ref{eq:fit_Qimp}) is shown by solid lines in panel (b) of Fig.~\ref{fig:Qimp_Tm}, while the absolute error with respect to the computed values of $Q_{\rm imp}$ in the MCP approach is displayed in panel (c).
We can see that the error remains relatively small, $|\Delta Q_{\rm imp}| \approx 5$ at most, in the whole inner-crust density range.
Indeed, even if the fit tends to break down at higher densities and higher temperatures (see black and blue lines in panel (b) of Fig.~\ref{fig:Qimp_Tm} for $n_B \gtrsim 0.05$~fm$^{-3}$), the error is very small because of the low values of $Q_{\rm imp}\approx 0.1-0.2$.

\begin{table}
\caption{Impurity parameter $Q_{\rm imp}$ in the  inner crust for different temperatures. The results are obtained with the fully self-consistent MCP calculation. }
\label{tab:Qimp}
\centering
\setlength{\tabcolsep}{5pt}
\begin{tabular}{l|ccccc}
\hline
\hline
\noalign{\smallskip}
 Density & \multicolumn{5}{c}{Temperature (MeV)}   \\  
 (fm$^{-3}$) & 0.5 & $k_{\rm B} T_{\rm m}$ &  1.0 & 1.5 & 2.0 \\
\noalign{\smallskip} \hline 
 \hline
 \noalign{\smallskip}
0.001   & 6.73  & 4.00  & 13.41 & 21.52 & 55.75 \\
0.005   &8.67   &7.31   &18.16  &49.65  &106.76\\
0.010   &10.74  &10.88  &24.69  &117.81 &23.21\\
0.015   &12.85  &14.42  &41.62  &81.21  &3.06\\
0.020   &15.02  &18.27  &84.15  &21.20  &1.08\\
0.025   &17.68  &23.58  &114.79 &4.41   &0.61\\
0.030   &20.89  &36.22  &77.58  &1.57   &0.41\\
0.035   &26.62  &70.06  &29.61  &0.84   &0.30\\
0.040   &40.09  &108.84 &8.26   &0.54   &0.23\\
0.045   &71.04  &81.13  &2.67   &0.39   &0.19\\
0.050   &108.29 &25.97  &1.25   &0.29   &0.16\\
0.055   &89.51  &4.76   &0.74   &0.24   &0.14\\
0.060   &33.78  &1.28   &0.52   &0.21   &0.14\\
0.065   &7.29   &0.57   &0.41   &0.20   &0.16\\
0.070   &2.27   &0.35   &0.39   &0.24   &0.19\\
\noalign{\smallskip}
\hline 
 \hline
\end{tabular}
\end{table}

\begin{table}
        \caption{Fitting parameters $a_{kj}$ in Eq.~(\ref{eq:fit_Qimp}) for the impurity parameter.}
        \label{tab:Qimp_fit_params}
        \centering
        \setlength{\tabcolsep}{5pt}
        \begin{tabular}{l|ccc}
                \hline
                \hline
                \noalign{\smallskip}
                 $a_{kj}$& & &  \\  
                  & $j = 1$ & $j  = 2$ &  $j = 3$\\
                \noalign{\smallskip} \hline 
                \hline
                \noalign{\smallskip}
                $k = 0$ & $  -2.285 \times 10^1$ & 
        $2.205 \times 10^1$ & $-5.370 $  \\
                $k = 1$ &$1.818\times 10^3$  & $ -1.087 \times 10^3$ & $2.000\times 10^2$\\
                $k = 2$ &$-9.456 \times 10^4$ &  $ 6.849\times 10^4$ & $ -2.673  \times 10^4$   \\
                $k = 3$ & $1.730 \times 10^6$  & $  -5.766\times 10^5$ & $4.969 \times 10^5$     \\
                $k = 4$ &$-1.161 \times 10^7$ &  $ 6.356 \times 10^6$ & $ -1.666 \times 10^7$ \\
                \noalign{\smallskip}
                \hline 
                \hline
        \end{tabular}
\end{table}

\section{Conclusions}
\label{sec:conlusions}

In this work, we studied the properties of the beta-equilibrated inner crust of a PNS in the liquid phase within a multi-component approach. 
To this aim, the formalism of \citet{Fantina2020} and \citet{Carreau2020b} was extended for its applicability over a wider range of densities and temperatures in the inner crust, from the neutron drip up to the crust--core transition. 
The cluster energetics was described within a CLD model approach, in which the surface parameters were optimised to reproduce the experimental masses from the AME2016 \citep{AME2016}, and the nuclear matter properties were calculated within finite-temperature mean-field thermodynamics. 
We performed the calculations employing both a perturbative MCP treatment, where the chemical potentials and gas densities were taken from the OCP calculations, and a full self-consistent MCP approach.

We find that, if non-linear mixing terms arising from the centre-of-mass motion of the ion are neglected, the most probable clusters in the MCP (perturbative) approach coincide with the OCP predictions.
However, as already discussed in \citet{Dinh2022}, this translational free-energy term should be taken into account in the calculations of the (liquid) inner-crust composition, leading to a breakdown in the ensemble equivalence.
Our outcomes also show that the OCP and the perturbative MCP treatments are a good approximation at relatively low densities and temperatures, especially as far as global properties such as the equation of state are concerned.
However, we believe that a full self-consistent MCP is needed for reliable predictions of the PNS crust composition, particularly in the deeper region of the crust.

Moreover, our findings reveal that, with increasing density and temperature, the abundance of light nuclei becomes important, and eventually dominates the whole distribution. 
This result has an important effect on the impurity parameter, and therefore potentially on the NS cooling. 
For applications to numerical simulations, we also provide both values of the impurity parameter calculated within our fully self-consistent MCP approach in tabular form and a fitting formula able to reproduce the numerical results with good accuracy.

In our study, we only considered spherical clusters. However, non-spherical structures, so-called pasta phases, may appear at the bottom of the inner crust (see \citet{Pelicer2021} for recent calculations of pasta phases in a MCP approach at finite temperature within a relativistic mean-field functional).
We defer an investigation of the co-existence of different (non-spherical) structures within our approach to future studies.

\begin{acknowledgements}
This work has been partially supported by the IN2P3 Master Project NewMAC, the ANR project `Gravitational waves from hot neutron stars and properties of ultra-dense matter' (GW-HNS, ANR-22-CE31-0001-01), and the CNRS International Research Project (IRP) `Origine des \'el\'ements lourds dans l’univers: Astres Compacts et Nucl\'eosynth\`ese (ACNu)'.  
\end{acknowledgements}

\begin{appendix}


\section{Pressure in the MCP}
\label{sec:Ptot}

The pressure in each WS cell can be written as 
\begin{equation}
        P^\j = P_i^\j + P_{{\rm g}n}  + P_e  \ ,
        \label{eq:POCP}
\end{equation}
where the gas and electron pressures are the same in all cells since both gases are uniformly distributed over the whole volume, and $P_i^\j$ is the ion pressure, 
\begin{equation}
        P_i^\j = P_{\rm trans}^{\star,\j} + P_{\rm int}^\j \ .
        \label{eq:Pij}
\end{equation}
The first term in Eq.~(\ref{eq:Pij}), $P_{\rm trans}^{\star,\j}$ , accounts for the contribution from the translational centre-of-ion motion:
\begin{equation}
        P_{\rm trans}^{\star,\j} = P_{\rm trans}^{\star} = \frac{k_{\rm B} T}{{\Bar{u}_{\rm f}} \langle V_{\rm WS} \rangle} \ ,
\end{equation}
while the second term, $P_{\rm int}^{(j)}$, is the interaction (Coulomb) term of the pressure as calculated in the (pure phase) OCP approximation (see also Eq.~(\ref{eq:P_int})).

To derive the total pressure in the MCP, we used the definition of the pressure in the canonical ensemble, 
\begin{equation}
    P^{\rm MCP} = - \left.\frac{\partial F}{\partial V}\right|_{{p_j},T} = 
    P_{{\rm g}n}+ P_e  +  P_i  \ ,
    \label{eqMCP:pressure_1}
\end{equation}
where 
$P_{{\rm g}n} = \mu_{{\rm g}n} \nng - \mathcal{F}_{\rm g}$
($P_e = \mu_e n_e - \mathcal{F}_e$) is the neutron gas\footnote{{Here, we use the notation $P_{{\rm g}n}$ to indicate the pressure of the self-interacting neutron gas, while in the main text $P_{\rm g}$ includes in-medium effects.}} (electron) pressure, $P_i$ is the ion pressure, ${p_j}$ is the set of probabilities of the different species, with $p_j$ being the probability of each component $j$.
However, as already pointed out by \citet{Fantina2020}, the total pressure in MCP is not equal to the sum of the pressure of the OCP phases, that is, $P^{\rm MCP} \neq \sum_j p_j P^{(j)}$, but it is rather given by
\begin{equation}
    P^{\rm MCP} =  
    P_{{\rm g}n} + P_e  +  P_{\rm trans}^{\star} + \frac{n_p^2}{\langle Z \rangle} \sum_j p_j \frac{\partial F^\j_{\rm Coul}}{\partial n_p} \ .
    \label{eqMCP:pressure_1}
\end{equation}

\end{appendix}


\begin{thebibliography}{}

\bibitem[Avancini et al.~(2009)]{Avan2009} Avancini, S.~S., Brito, L., Marinelli, J.~R., et al.\ 2009, Phys. Rev. C, 79, 035804
\bibitem[Avancini et al.~(2017)]{Avan2017} Avancini, S.~S., Ferreira, M., Pais, H., Provid\^{e}ncia, C., \& R\"{o}pke, G.\ 2017, Phys. Rev. C, 95, 045804

\bibitem[Barros et al.~(2020)]{Barros2020} Barros, C.~C., Menezes, D.~P., \& Gulminelli, F. \ 2020, Phys. Rev. C, 101, 035211
\bibitem[Baym et al.~(1971)]{bbp} Baym, G., Bethe, H.~A., \& Pethick, C.~J.\ 1971, Nucl. Phys. A, 175, 225
\bibitem[Botvina \& Mishustin~(2010)]{Botvina2010} Botvina, A.~S., \&  Mishustin, I.~N.\ 2010, Nucl. Phys. A, 843, 98

\bibitem[Burrows \& Lattimer~(1984)]{Burrows1984} Burrows, A., \& Lattimer, J.~M.\ 1984, \apj, 285, 294

\bibitem[Caplan et al.~(2021)]{Caplan2021} Caplan, M. E., Forsman, C. R., \&  Schneider, A. S. \ 2021, Phys. Rev. C, 103, 055810

\bibitem[Carreau et al.~(2020a)]{Carreau2020a} Carreau, T., Gulminelli, F., Chamel, N., Fantina, A.~F., \&  Pearson, J.~M.\ 2020a, A\&A, 635, A84  

\bibitem[Carreau et al.~(2020b)]{Carreau2020b} Carreau, T., Fantina, A.~F., \& Gulminelli, F.\ 2020b, A\&A 640, A77; Carreau, T., Fantina, A.~F., \& Gulminelli, F.\ 2021, A\&A 645, C1

\bibitem[Dinh Thi et al.~(2021)]{Dinh2021a} Dinh Thi, H., Carreau, T., Fantina, A.~F., \& Gulminelli, F.\ 2021, A\&A, 654, A114

\bibitem[Dinh Thi et al.~(2023)]{Dinh2022} Dinh Thi, H., Fantina, A. F., \& Gulminelli, F.\ 2023, A\&A, 672, A160

\bibitem[Ducoin et al.~(2007)]{Ducoin2007} Ducoin, C., Chomaz, Ph., \& Gulminelli, F.\ 2007, Nucl. Phys. A, 789, 403

\bibitem[Fantina et al.~(2020)]{Fantina2020} Fantina, A.~F., De Ridder, S., Chamel, N., \& Gulminelli, F.\ 2020, A\&A 633, A149

\bibitem[Furusawa et al.~(2017)]{Furusawa2017} Furusawa, S., Togashi, H., Nagakura, H., et al.\ 2017, J. Phys. G, 44, 094001

\bibitem[Goriely et al.~(2011)]{Goriely2011} Goriely, S., Chamel, N., Janka, H.-Th., \& Pearson, J.~M.\ 2011, A\&A, 531, A78

\bibitem[Goriely et al.~(2013)]{BSK24} Goriely, S., Chamel, N., \& Pearson, J.~M.\ 2013, Phys. Rev. C, 88, 024308

\bibitem[Gourgouliatos \& Esposito~(2018)]{Gour2018} Gourgouliatos, K.~N., \& Esposito, P.\ 2018, in The Physics and Astrophysics of Neutron Stars, eds. L. Rezzolla, P. Pizzochero, D.~I. Jones, N. Rea, I. Vidaña (Cham: Springer), Astrophys. Space Sci. Lib., 457, 57

\bibitem[Grams et al.~(2018)]{Grams2018} Grams, G., Giraud, S., Fantina, A.~F., \& Gulminelli, F.\ 2018, Phys. Rev. C, 97, 035807

\bibitem[Gulminelli \& Raduta~(2015)]{gulrad2015} Gulminelli, F., \& Raduta, Ad.~R.\ 2015, Phys. Rev. C, 92, 055803

\bibitem[Haensel et al.~(2007)]{hpy2007} Haensel, P., Potekhin, A.~Y., \& Yakovlev, D.~G.\ 2007, ``Neutron Stars 1. Equation of state and structure'' (New York: Springer)

\bibitem[Hempel \& Schaffner-Bielich~(2010)]{Hempel2010} Hempel, M., \& Schaffner-Bielich, J.\ 2010, Nucl. Phys. A, 837, 210

\bibitem[Horowitz et al.~(2015)]{Horowitz2015} Horowitz, C. J., Berry, D. K., Briggs, C. M., et al.\ 2015, Phys. Rev. Lett., 114, 031102

\bibitem[Jones~(2004)]{Jones2004} Jones, P.~B.\ 2004, Phys. Rev. Lett., 93, 221101

\bibitem[Lattimer \& Swesty~(1991)]{Lattimer1991} Lattimer, J.~M., \& Swesty, F.~D.\ 1991, Nucl. Phys. A, 535, 331

\bibitem[Lattimer et al.~(1985)]{lattimer1985} Lattimer, J.~M., Pethick, C.~J., Ravenhall, D.~G., \& Lamb, D.~Q.\ 1985, Nucl. Phys. A, 432, 646

\bibitem[Margueron et al.~(2018a)]{Margueron2018a} Margueron, J., Hoffmann Casali, R., \& Gulminelli, F.\ 2018a, Phys. Rev. C, 97, 025805

\bibitem[Margueron et al.~(2018b)]{Margueron2018b} Margueron, J., Hoffmann Casali, R., \& Gulminelli, F.\ 2018b, Phys. Rev. C, 97, 025806

\bibitem[Maruyama et al.~(2005)]{Maru2005} Maruyama, T.,  Tatsumi, T., Voskresensky, D.~N.,  Tanigawa, T., \& Chiba, S.\ 2005, Phys. Rev. C, 72, 015802

\bibitem[Medin \& Cumming~(2010)]{Medin2010} Medin, Z., \& Cumming, A.\ 2010, Phys. Rev. E, 81, 036107

\bibitem[Newton ~(2013)]{Newton2013b} Newton, W. G.\ 2013, Nat. Phys., 9, 396

\bibitem[Newton et al.~(2013)]{Newton2013} Newton, W.~G., Gearheart, M., \& Li, B.-A.\ 2013, \apjs, 204, 9

\bibitem[Oertel et al.~(2017)]{Oertel2017} Oertel, M., Hempel, M., Kl\"{a}hn, T., \& Typel, S.\ 2017, Rev. Mod. Phys., 89, 015007

\bibitem[Onsi et al.~(2008)]{Onsi2008} Onsi, M., Dutta, A.~K., Chatri, H., et al.\ 2008, Phys. Rev. C, 77, 065805

\bibitem[Page et al.~(2006)]{Page2006} Page, D., Geppert, U., \& Weber, F.\ 2006, Nucl. Phys. A, 777, 497

\bibitem[Pais et al.~(2018)]{Pais2018} Pais, H., Gulminelli, F., Provid\^{e}ncia, C., \& R\"{o}pke, G.\ 2018, Phys. Rev. C, 97, 045805

\bibitem[Pelicer et al.~(2021)]{Pelicer2021} Pelicer, M.~R., Menezes, D.~P., Barros, C.~C., \& Gulminelli, F.\ 2021, Phys. Rev. C, 104, L022801

\bibitem[Pons et al~(1999)]{Pons1999} Pons, J.~A., Reddy, S., Prakash, M., Lattimer, J.~M., \& Miralles, J.~A.\ 1999, \apj, 513, 780

\bibitem[Pons et al.~(2013)]{Pons2013} Pons, J.~A., Vigan\`o, D., \& Rea, N.\ 2013, Nat. Phys., 9, 431

\bibitem[Prakash et al.~(1997)]{Prakash1997} Prakash, M., Bombaci, I., Prakash, M., et al.\ 1997, Phys. Rep., 280, 1

\bibitem[Raduta \& Gulminelli~(2019)]{radgul2019} Raduta, Ad.~R., \& Gulminelli, F.\ 2019, Nucl. Phys. A, 983, 252

\bibitem[Ravenhall et al.~(1983)]{Ravenhall1983} Ravenhall, D.~G., Pethick, C.~J., \& Lattimer, J.~M.\ 1983, Nucl. Phys. A, 407, 571

\bibitem[R\"{o}pke~(2020)]{Roepke2020} R\"{o}pke, G.\ 2020, Phys. Rev. C, 101, 064310

\bibitem[Schmitt \& Shternin~(2018)]{Schmitt2018} Schmitt, A., \& Shternin, P.\ 2018, in The Physics and Astrophysics of Neutron Stars, eds. L. Rezzolla, P. Pizzochero, D.~I. Jones, N. Rea, I. Vidaña (Cham: Springer), Astrophys. Space Sci. Lib., 457, 455

\bibitem[Schneider et al.~(2016)]{Schneider2016} Schneider, A. S., Berry, D. K., Caplan, M. E., Horowitz, C. J., \& Lin, Z.\ 2016, Phys. Rev. C, 93, 065806

\bibitem[Souza et al.~(2009)]{Souza2009} Souza, S.~R., Steiner, A.~W., Lynch, W.~G., Donangelo, R., \& Famiano, M.~A.\ 2009, ApJ, 707, 1495

\bibitem[Vigan\`o et al.~(2013)]{Vigano2013} Vigan\`o, D., Rea, N., Pons, J.~A., et al.\ 2013, MNRAS, 434, 123

\bibitem[Wang et al.~(2017)]{AME2016} Wang M., Audi G., Kondev F.~G., et al.\ 2017, Chinese Phys. C, 41, 030003 

\bibitem[Yakovlev \& Pethick~(2004)]{Yakovlev2004} Yakovlev, D.~G., \& Pethick, C.~J.\ 2004, Annu. Rev. Astron. Astrophys., 42, 169


\end{thebibliography}
\end{document}